\documentclass[12pt]{iopart}
\usepackage{iopams}
\usepackage{graphicx}
\usepackage{amssymb}
\usepackage{bm}
\usepackage{siunitx}
\usepackage[colorlinks=true,linkcolor=blue,citecolor=blue,urlcolor=blue]{hyperref}

\begin{document}

\title[Interior contacts in a narrow quantum Hall bar]{Interior contacts in a narrow quantum Hall bar: a two-dimensional self-consistent screening calculation of the current distribution}

\author{Afif Siddiki}
\address{Atlas University Vocational School, \.Istanbul, Turkey}
\ead{afif.siddiki@atlas.edu.tr}

\begin{abstract}
We present fully two-dimensional self-consistent calculations of the electron density, the local filling factor and the current distribution in a six-terminal GaAs/AlGaAs Hall bar of width $W=3\,\mu$m with two additional square contacts, A and B, placed on the axis of the channel and connected to the outside by air bridges, following the interior-contact experiment of Kendirlik \emph{et al.} [Nat.\ Commun.\ \textbf{8}, 14082 (2017)] on a $10\,\mu$m wide bar. Within the screening theory of the integer quantized Hall effect, the Thomas--Fermi--Poisson equations are solved on the full device plane, including the probe arms and the depletion regions around the interior contacts, and the current is obtained from a local Ohm's law with a filling-factor dependent conductivity tensor, for source--drain and for interior-contact excitation separately. Across the $\nu=2$ plateau the interior contacts pass through three regimes as the field is lowered: galvanically connected to the current path in a compressible bulk, floating in the Hall potential of a fully incompressible bulk, and isolated from both the bulk and the current-carrying edge strips by closed incompressible rings that encircle each contact. The two-terminal resistance between the interior contacts rises by orders of magnitude in the latter two regimes, and the field interval of this isolation lies inside the Hall plateau and ends before it on the low-field side; its high-field end is sharply defined by the disappearance of the incompressible bulk, its low-field end shrinks with temperature and Landau-level broadening. The voltage between an interior contact and a perimeter probe approaches one half of the quantized Hall resistance across the plateau. The topology of the incompressible regions and the boundaries of the isolation interval are shown to be insensitive to the regularization of the local conductivity, to the grid and to moderate disorder, whereas the magnitude of the isolation resistance is not.
\end{abstract}


\submitto{\JPCM}
\maketitle

\section{Introduction}
\label{sec:intro}

The question of where the current flows in a two-dimensional electron system (2DES) under quantized Hall conditions~\cite{Klitzing80,Tsui82} has been debated since the discovery of the effect. In the single-particle picture~\cite{Laughlin81,Halperin82,Buttiker86,Buttiker88} the bulk is an insulator with the Fermi energy in a Landau gap, and the current is carried by one-dimensional channels at the sample boundary; this picture underlies the topological description of the Hall conductance as a Chern number~\cite{Thouless82,Kohmoto85,HasanKane10}, which is defined for a translation-invariant bulk and whose connection to the current distribution in a finite, contacted device is not direct. The self-consistent treatment of the electrostatics~\cite{Chang90,Chklovskii92,Chklovskii93,Lier94,Oh97}, and in particular the screening theory of the integer quantized Hall effect (IQHE)~\cite{Guven03,Siddiki04,SiddikiGerhardts03}, replaces the edge channels by incompressible strips of finite width that host a dissipationless Hall current, and shows that the width and even the existence of these strips depend on the filling factor, the confinement, the temperature and the Landau-level broadening. In this description the bulk of a Hall bar is compressible for most of the plateau interval and is incompressible only in a subinterval near its high-field end. Scanning-probe imaging of the Hall potential~\cite{Weitz00,Ahlswede01,Ahlswede02,Dahlem10,Suddards12} and single-electron-transistor measurements of the local compressibility~\cite{Ilani04} agree with this picture: near the low-field side of a plateau the potential drops across narrow strips at the edges, near the high-field side it drops nonlinearly across a wide central region, and outside the plateau it varies linearly across the sample. Current-induced asymmetries of the strips~\cite{Siddiki09,Siddiki10,Gerhardts13} and the influence of the contacts~\cite{Eksi10,Dahlem10} have been treated within the same framework, and more recent quantum-transport and density-functional calculations~\cite{Champel08,Oswald17,Armagnat20,Kloss21} reach compatible conclusions.

An alternative, purely transport-based way of probing the interior of a Hall bar is to place ohmic contacts \emph{inside} the channel and to connect them to the outside by air bridges~\cite{Yildiz14,Kendirlik17,Sirt23,Sirt25}; with such a device the transition from edge- to bulk-currents across a plateau has recently been demonstrated directly~\cite{Sirt25}. Such contacts change the genus of the sample in coordinate space without disturbing the boundary of the 2DES. A measurement of the voltage between two interior contacts, $V_{AB}$, while a small current is driven between them, gives direct information on whether the interior of the sample is connected to them or isolated from them: the impedance between the contacts contains, besides an ohmic part, a quantum capacitance proportional to the thermodynamic density of states at the Fermi energy, which vanishes wherever the 2DES is incompressible~\cite{Yildiz14}. On a $10\,\mu$m wide bar it was found that $V_{AB}$ saturates at a constant value only in a subinterval of each Hall plateau, that this subinterval is not centred on the plateau, and that it shrinks faster with temperature than the plateau itself~\cite{Kendirlik17}. These observations were interpreted in terms of the screening theory: at the high-field side of the saturation interval the contacts are decoupled by an incompressible bulk, at its low-field side by incompressible rings that encircle each contact while the bulk between them is already compressible, a situation reminiscent of the Corbino geometry~\cite{Dolgopolov92,Liu98,Woltjer86}. The accompanying calculations~\cite{Yildiz14,Sirt23} were performed for wider bars ($7$--$8\,\mu$m), with periodic boundary conditions, ideal source and drain contacts, without probe arms, and for a few selected magnetic fields.

In the present work we address the interior-contact geometry by a fully two-dimensional self-consistent calculation for a much narrower bar, $W=3\,\mu$m, that includes the actual six-terminal layout with its probe arms, the finite depth of the interior contacts and the doping structure. In such a bar the interior contacts occupy a sizeable fraction of the channel, and the electrostatic interaction between the contacts, the arms and the channel edges can no longer be neglected. The choice of a narrow bar is deliberate: the width of the incompressible strips at the edges is set by the electrostatics and is of the order of $\SI{100}{nm}$, so that in a $3\,\mu$m channel the strips, the depletion regions around the interior contacts and the bulk are all resolved on the same footing, and the transition from bulk to edge transport across the plateau happens within a narrow interval of the magnetic field. Narrow, high-mobility bars were also the samples recommended in Ref.~\cite{Yildiz14} for a clean test of bulk compressibility.

The paper is organized as follows. Section~\ref{sec:model} describes the device geometry, the measurement circuit and the two-dimensional Thomas--Fermi--Poisson scheme together with the local current model. Section~\ref{sec:results} presents the zero-field density, the filling-factor landscape at $\nu_{\rm nom}=2.1$ and the current and Hall-potential distributions at five filling factors across the $\nu=2$ plateau. Section~\ref{sec:discussion} gives the field sweeps of the source--drain and interior-contact response coefficients, the tests of their robustness against the model parameters, the grid, disorder and contact position, and the comparison with the interior-contact experiments; Sec.~\ref{sec:conclusion} summarizes.

\section{Device and model}
\label{sec:model}

\subsection{Geometry and measurement scheme}
\label{sec:geometry}

\begin{figure}
\includegraphics[width=0.9\textwidth]{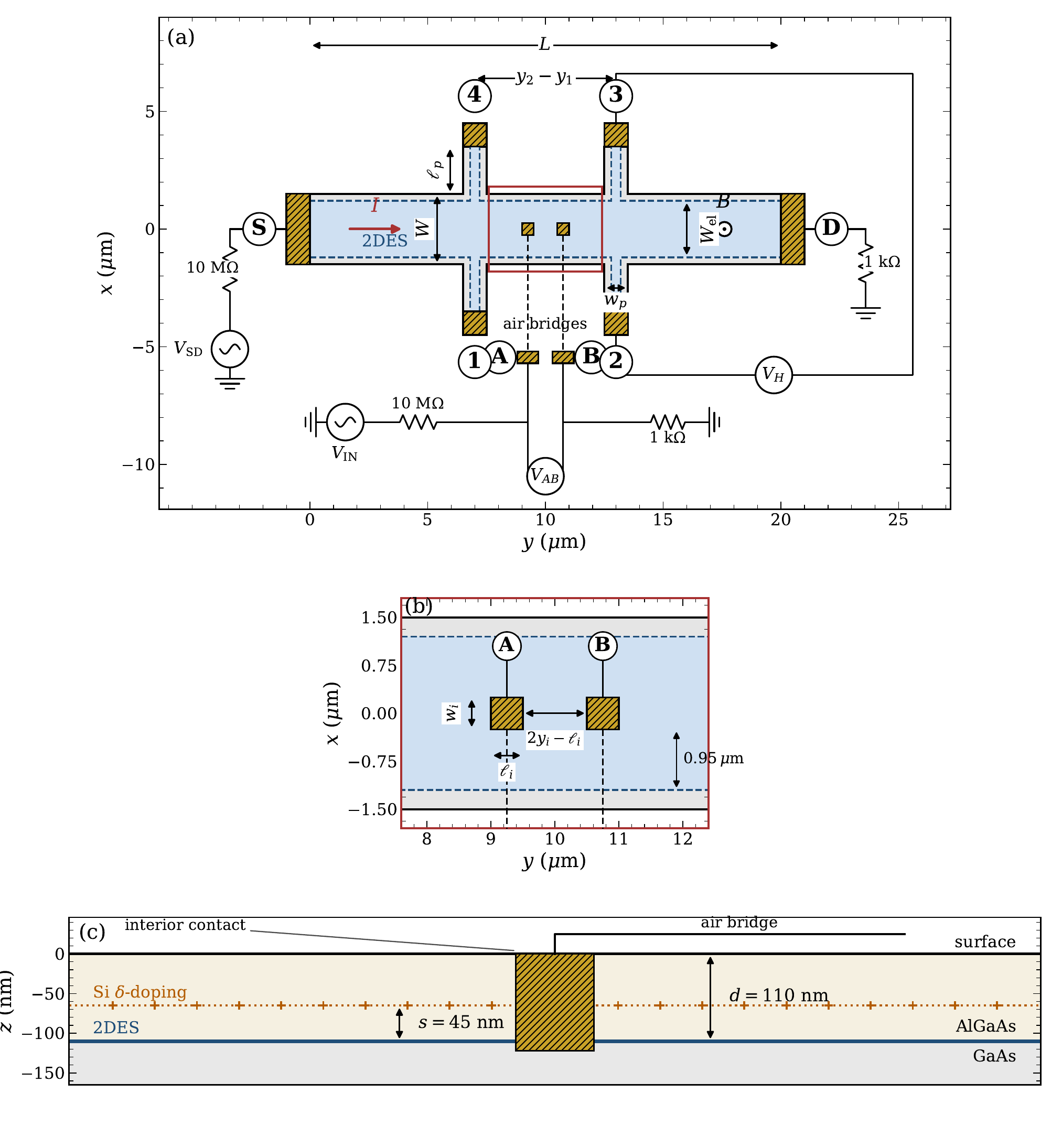}
\caption{(a) Layout of the six-terminal Hall bar (width $W=3\,\mu$m, length $L\approx 20\,\mu$m) with source S, drain D, perimeter probes 1--4 and the two interior contacts A and B, together with the measurement circuit. The source--drain current $I$ is set by $V_{\rm SD}$ through a $\SI{10}{M\ohm}$ resistor; a second current can be injected between A and B by $V_{\rm IN}$. $V_H$ is measured between probes 2 and 3 and $V_{AB}$ between the interior contacts, which are connected by air bridges. The dashed lines mark the electronic edge of the 2DES. (b) Zoom onto the interior contacts: each is a $\ell_i\times w_i=0.5\times0.5\,\mu$m$^2$ square, the gap between them is $2y_i-\ell_i=1\,\mu$m, and their distance to the electronic edge is $0.95\,\mu$m. (c) Cross-section of the heterostructure: the 2DES lies $d=\SI{110}{nm}$ below the surface, the Si $\delta$-doping $s=\SI{45}{nm}$ above it, and the interior contact reaches down to the 2DES plane.}
\label{fig:geometry}
\end{figure}

Figure~\ref{fig:geometry}(a) shows the layout: a six-terminal Hall bar with two additional interior contacts, i.e. eight contacts in total. A Hall bar of length $L\approx 20\,\mu$m and lithographic width $W=3\,\mu$m is defined in a GaAs/AlGaAs heterostructure and is contacted at its ends by the source (S) and drain (D). Four perimeter probes, 1--4, are attached through arms of width $w_p\approx 0.5\,\mu$m and length $\ell_p\approx 1.5\,\mu$m, positioned pairwise at $y_1\approx 7\,\mu$m and $y_2\approx 13\,\mu$m, so that $y_2-y_1\approx 6\,\mu$m. The Hall voltage $V_H$ is measured between contacts 2 and 3 and the longitudinal voltage between 1 and 2 or 3 and 4.

Two square interior contacts, A and B, are placed on the axis of the bar ($x=0$) at $y_A\approx 9.25\,\mu$m and $y_B\approx 10.75\,\mu$m. Each has a side length $\ell_i=w_i=0.5\,\mu$m, so that the metallic edges of A and B are $1\,\mu$m apart and each contact is $0.95\,\mu$m away from the electronic edge of the channel [Fig.~\ref{fig:geometry}(b)]. The contacts are connected to bond pads outside the mesa by air bridges, which cross the 2DES without touching it~\cite{Kendirlik17}. Compared with the $10\,\mu$m bar of Ref.~\cite{Kendirlik17}, where the contacts had an area of about $1\,\mu{\rm m}^2$ and were $7.5\,\mu$m apart, both the contacts and their separation are scaled down.

The excitation scheme follows Refs.~\cite{Kendirlik17,Sirt23}. A source voltage $V_{\rm SD}$ is applied through a $\SI{10}{M\ohm}$ series resistor, so that S--D is current biased ($I\approx V_{\rm SD}/\SI{10}{M\ohm}$), and the drain is grounded through a $\SI{1}{k\ohm}$ sensing resistor. Independently, a second current can be driven between A and B by $V_{\rm IN}$ through an identical $\SI{10}{M\ohm}$--$\SI{1}{k\ohm}$ pair, and the voltage $V_{AB}$ between the interior contacts is recorded. In the experiment the two excitations are applied at different lock-in frequencies so that both signals can be separated~\cite{Kendirlik17}. The two circuits can be used separately (S--D current, A--B as a voltage probe; or A--B current, perimeter probes as voltage probes) or simultaneously.

The vertical structure is sketched in Fig.~\ref{fig:geometry}(c). The 2DES resides at a depth $d=\SI{110}{nm}$ below the surface at the GaAs/AlGaAs interface; a Si $\delta$-doping layer lies $s=\SI{45}{nm}$ above the 2DES. The interior contacts are modelled as metallic columns that penetrate the AlGaAs and reach the plane of the 2DES, where they are treated as equipotentials at zero field and as regions of vanishing electron density in the transport calculation. This is the ``low-resistance, reflective'' contact of Ref.~\cite{Eksi10}: the 2DES is depleted in front of the metal over a distance of the order of $\SI{100}{nm}$, in agreement with the scanning-probe observations of Dahlem \emph{et al.}~\cite{Dahlem10}, and the contact couples to the 2DES through the compressible rim of this depletion region.

\subsection{Two-dimensional Thomas--Fermi--Poisson scheme}
\label{sec:tfpa}

The electrostatics is treated on the full $(x,y)$ plane of the device. The electron density $n_{\rm el}(\bm r)$ and the total electrostatic potential energy $V(\bm r)=V_{\rm bg}(\bm r)+V_{\rm H}(\bm r)$ are determined self-consistently in the Thomas--Fermi--Poisson approximation (TFPA)~\cite{Lier94,Oh97,Siddiki04},
\begin{equation}
n_{\rm el}(\bm r)=\int dE\, D(E)\,f\!\left(\frac{E+V(\bm r)-\mu^\star}{k_{\rm B}T}\right),
\label{eq:tfa}
\end{equation}
where $D(E)$ is the density of states, $f$ the Fermi function, $\mu^\star$ the electrochemical potential and $T$ the temperature. The Hartree term
\begin{equation}
V_{\rm H}(\bm r)=\frac{e^2}{4\pi\epsilon_0\bar\kappa}\int d^2r'\,K(\bm r-\bm r')\,n_{\rm el}(\bm r')
\label{eq:hartree}
\end{equation}
is obtained from $n_{\rm el}$ with the kernel of the layered structure of Fig.~\ref{fig:geometry}(c), $K(\bm u)=1/|\bm u|-1/\sqrt{|\bm u|^2+(2d)^2}$, i.e. the bare Coulomb interaction in the plane of the 2DES together with its image in the surface, at which the Fermi level is pinned at midgap by surface states ($\bar\kappa=12.4$ for GaAs). The background $V_{\rm bg}$ is generated in the same way by the ionized donors of the $\delta$-layer at a distance $s$ above the 2DES, with the kernel $1/\sqrt{|\bm u|^2+s^2}-1/\sqrt{|\bm u|^2+(2d-s)^2}$. The donors are removed within $l_{\rm pin}=\SI{150}{nm}$ of every etched edge and of every contact window, which represents the compensation by surface states and the alloyed region of a contact and produces the steep, mesa-like depletion of Ref.~\cite{Arslan08}; their density, $n_D=5.20\times10^{15}\,{\rm m}^{-2}$, is chosen so that the electron density in the plain channel at $B=0$ is $n_0=3.0\times10^{15}\,{\rm m}^{-2}$. Both convolutions are evaluated by zero-padded fast Fourier transforms on a rectangular grid of $25\times25\,{\rm nm}^2$ cells covering the device and a margin of $0.2\,\mu$m on each side ($20.4\times7.4\,\mu{\rm m}^2$, $816\times296$ cells); the zero padding implements open boundary conditions without periodic images. The $1/|\bm u|$ part of the kernel is integrated analytically over each source cell. All depletion regions---of the arms, of the channel edges and of the interior contacts---are thus obtained from the same Poisson solution. The interior contacts are regions of vanishing electron density; they couple to the 2DES through the compressible rim of the depletion halo that forms in front of them (Sec.~\ref{sec:ohm}).

At zero magnetic field $D(E)=m^\ast/\pi\hbar^2$ is constant; at finite $B$ we use Gaussian-broadened Landau levels,
\begin{equation}
D(E)=\frac{2eB}{h}\sum_{n\ge0}\frac{1}{\sqrt{2\pi}\,\Gamma}\exp\!\left[-\frac{(E-E_n)^2}{2\Gamma^2}\right],\qquad E_n=\hbar\omega_c\left(n+\frac12\right),
\label{eq:dos}
\end{equation}
with a width $\Gamma=\SI{0.3}{meV}\ll\hbar\omega_c$, so that the local filling factor is $\nu(\bm r)=n_{\rm el}(\bm r)h/eB$. Spin degeneracy is retained ($g_s=2$); at the fields considered the bare Zeeman energy $g^\ast\mu_{\rm B}B\approx0.15$--$\SI{0.2}{meV}$ is comparable to $\Gamma$ but small compared with $\hbar\omega_c\approx\SI{10}{meV}$, and since the calculations focus on the spin-degenerate $\nu=2$ plateau, where the Fermi energy lies in the cyclotron gap, the spin splitting affects neither the position nor the width of the $\nu=2$ incompressible regions~\cite{Arslan08}; odd-integer strips would require the exchange-enhanced gap and are not considered. The temperature is $T=\SI{1.5}{K}$ throughout ($k_{\rm B}T/\hbar\omega_c\approx0.013$), low enough for the incompressible regions to be well developed~\cite{Oh97,Siddiki04}; the dependence on $T$ and $\Gamma$ is examined in Sec.~\ref{sec:robust}. The nonlinear problem $n_{\rm el}=n_{\rm TF}[\mu^\star-V(n_{\rm el})]$ is solved at each field by Anderson-accelerated Kerker mixing followed by a Newton--Krylov iteration with a Fourier-space preconditioner, to a residual $\max|n_{\rm el}-n_{\rm TF}|<10^{-6}n_0$; the fields of a sweep are solved in sequence, each starting from its neighbour.

The TFPA neglects the finite extent of the Landau wave functions and therefore produces incompressible strips also where they are narrower than the Fermi wavelength $\lambda_F=\sqrt{2\pi/n_0}\approx\SI{46}{nm}$, i.e. where they cannot support a dissipationless current~\cite{Siddiki04,SiddikiMarquardt07,Suzuki93}. The standard remedy is to average the conductivity tensor over a box of width $\lambda_F$ before solving the transport problem~\cite{Siddiki04}. On the $25\,$nm grid used here the cell is $\lambda_F/2$ wide, and we have checked on a $10\times25\,{\rm nm}^2$ grid with explicit averaging that the plateau results are the same (Sec.~\ref{sec:robust}); the coarse grid does not, however, replace the averaging where strips are only one or two cells wide, and the corresponding results (the rings at the low-field end of the plateau) are marked as upper bounds. The TFPA itself is used, rather than a Hartree or density-functional calculation with Landau-level wave functions, because the device contains $\sim2\times10^5$ cells and each field of a sweep has to be solved in seconds; for the quantities that enter here---the positions and widths of the incompressible strips---the coarse-grained TFPA has been benchmarked against Hartree~\cite{Siddiki04,SiddikiMarquardt07} and against quantum-transport and density-functional calculations~\cite{Champel08,Oswald17,Armagnat20,Kloss21}. It cannot describe tunnelling across a strip, which is discussed in Sec.~\ref{sec:limits}.

Cells in which $|\nu(\bm r)-2|<0.01$ and the local thermodynamic density of states satisfies $\partial n_{\rm el}/\partial\mu^\star<0.1\,D_0$, with $D_0=m^\ast/\pi\hbar^2$, are identified as incompressible; all other cells are compressible. Throughout, $\nu_{\rm bulk}$ denotes the filling factor at the centre of the plain channel, at $(x,y)=(0,4\,\mu{\rm m})$, and the fields are labelled by the nominal filling factor $\nu_{\rm nom}=n_0h/eB$ of the zero-field density; $\nu_{\rm bulk}$ is up to 1\% smaller than $\nu_{\rm nom}$ above the plateau, because the Landau quantization redistributes the electrons at fixed donor charge.

\subsection{Local Ohm's law}
\label{sec:ohm}

The current distribution is calculated from the local Ohm's law~\cite{Guven03,Siddiki04},
\begin{equation}
\bm j(\bm r)=\hat\sigma\big(\nu(\bm r)\big)\,\bm E(\bm r),\qquad \bm E=\nabla\mu^\star/e,\qquad \nabla\cdot\bm j=0,
\label{eq:ohm}
\end{equation}
with a local conductivity tensor built from
\begin{equation}
\sigma_\ell(\nu)=\frac{e^2}{h}\Big[\epsilon+\frac{\kappa\nu}{2}\big(1-\cos2\pi\nu\big)\Big],\qquad
\sigma_H(\nu)=\nu\frac{e^2}{h}.
\label{eq:sigma}
\end{equation}
Equation~(\ref{eq:sigma}) is a phenomenological interpolation, not a microscopic result: it reproduces the two limits that matter for the local Ohm's law, $\sigma_\ell\to0$ in incompressible regions (even $\nu$) and $\sigma_\ell$ of order $\nu e^2/h$ in compressible ones, with $\kappa=0.5$ fixing the compressible-bulk value to the magnitude found in the self-consistent Born approximation at half-integer filling~\cite{Ando82,Siddiki04}. The residual $\epsilon$ is a phenomenological proxy for all processes that give an incompressible region a finite conductance and that the Thomas--Fermi model does not contain: the disorder-induced density of states in the Landau-level tails, thermal activation across the gap, and tunnelling or scattering across a narrow strip. It regularizes the incompressible regions: the longitudinal resistivity $\rho_\ell=\sigma_\ell/(\sigma_\ell^2+\sigma_H^2)$ vanishes there as $\epsilon\to0$, so that these regions carry a dissipationless Hall current. Two consequences of this choice are kept in mind throughout. First, the geometry of the incompressible regions is fixed by the density solution of Sec.~\ref{sec:tfpa} and does not depend on Eq.~(\ref{eq:sigma}); second, any resistance that is determined by transport \emph{across} an incompressible region---the longitudinal resistance on a plateau, the two-terminal resistance between contacts separated by an incompressible ring---is proportional to $1/\epsilon$ (or $\epsilon$) and is therefore only meaningful up to this regularization. We report such quantities as ratios or as bounds and verify in Sec.~\ref{sec:robust} which results are independent of $\epsilon$ and $\kappa$; there we also show that the measured saturation of $V_{AB}$ fixes the magnitude of $\epsilon$, and estimate the width below which tunnelling makes a strip transparent, which a constant $\epsilon$ cannot describe.

\begin{figure}
\includegraphics[width=0.9\textwidth]{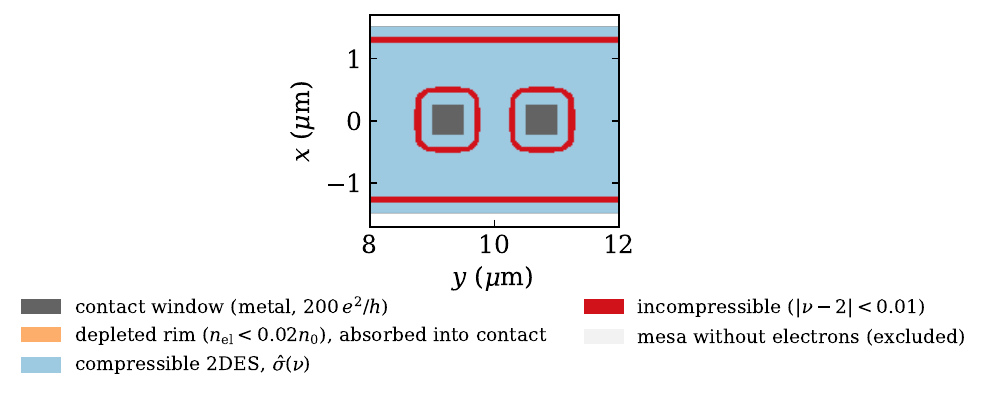}
\caption{Computational domains of the transport calculation around the interior contacts at $\nu_{\rm nom}=2.1$: contact windows (dark grey) and the depleted rims absorbed into them (orange) form the contact domain of conductivity $200\,e^2/h$; the compressible 2DES (blue) carries the local $\hat\sigma(\nu)$ of Eq.~(\ref{eq:sigma}); incompressible cells (red) have $\sigma_\ell=\epsilon e^2/h$.}
\label{fig:domains}
\end{figure}

Equation~(\ref{eq:ohm}) is discretized by a finite-volume scheme on the grid of the density calculation. The dissipative flux through a cell face is $\sigma_\ell$ times the potential difference of the two cells; the Hall flux through a face is $\sigma_H$ times the potential difference of the two corners of the face, corner potentials being the mean of the adjacent cells, so that the Hall fluxes around a cell telescope and the scheme is exactly divergence free for uniform $\sigma_H$. The computational domains are shown in Fig.~\ref{fig:domains}: (i) the contact windows (S, D, 1--4, A, B) are isotropic conductors of conductivity $200\,e^2/h$; (ii) the depleted rim in front of each window, where the self-consistent density is below $0.02\,n_0$ over a distance of up to $\SI{120}{nm}$, is absorbed into the contact domain, since cells without electrons cannot carry a current in the local model and would otherwise disconnect the contact; (iii) all other cells with $n_{\rm el}>0.02\,n_0$ are 2DES with the local $\hat\sigma(\nu)$. The metal--2DES interface is thus the boundary between domains (ii) and (iii), across which the same flux expressions apply; no interface resistance is included, i.e. the contacts are ideal ohmic contacts. The total current $I$ is injected uniformly over the cells of the source, the drain is grounded (one cell fixed at zero, the rest of the domain following from continuity), and every other contact is floating, i.e. carries no net current; the potential of a floating contact is the mean over its domain, which is uniform to $10^{-6}$ relative because of the high conductivity. Alternatively a current $I_{AB}$ is injected at A and extracted at B with S, D and 1--4 floating (A--B excitation). In linear response the two excitations are independent and each defines a set of transfer coefficients $V_i/I_j$; they are computed and reported separately. The linear system is solved by a sparse LU factorization; for $I=\SI{1}{nA}$ the net current into each floating contact is below $10^{-20}\,$A, the source and drain currents balance to $10^{-12}$ relative, and the largest divergence residual in any 2DES cell is $7\times10^{-22}\,$A. The imposed current is small enough ($I=\SI{1}{nA}$, Hall voltage $\SI{13}{\mu V}\ll\hbar\omega_c/e$) that the induced Hall potential does not modify the density distribution, i.e. we remain in the linear-response regime~\cite{Siddiki04,Arslan08}; the nonlinear regime, in which the current itself shifts the strips~\cite{Siddiki09,Siddiki10,Gerhardts13}, and local heating at the injection points~\cite{Akera00,Akera01,Kanamaru06,Yurdasan13,Eksi10} are discussed in Sec.~\ref{sec:limits}.

For an ac excitation at angular frequency $\omega$ the continuity equation acquires a displacement term, $\nabla\cdot[\hat\sigma\nabla\phi]-i\omega C_{\rm loc}\phi=0$ with $\phi=\mu^\star/e$, where the local capacitance per unit area, $1/C_{\rm loc}=d/\bar\kappa\epsilon_0+1/e^2(\partial n_{\rm el}/\partial\mu^\star)$, is the series combination of the geometric capacitance of the 2DES to the surface---the same pinned surface that provides the image plane of Eq.~(\ref{eq:hartree})---and the local quantum capacitance, which vanishes in the incompressible regions~\cite{Yildiz14}. This is a device impedance: the air bridges, bond pads, cables and the $\SI{10}{M\ohm}$ series resistors of the measurement circuit are not included, and the dc conductivity tensor is used at all frequencies, which restricts the model to $\omega\ll\omega_c$. The ac calculation is used only for the qualitative question of where in the plateau the interior-contact impedance acquires a capacitive phase (Sec.~\ref{sec:robust}).

Each of Figs.~\ref{fig:cur18}--\ref{fig:cur22} below shows, for one field, (a) the local filling factor $\nu(\bm r)$ with the incompressible cells in red, (b) the local longitudinal conductivity $\sigma_{xx}(\bm r)$ on a logarithmic scale, (c) the current density normalized to the total current, $|\bm j|/I$ (units $\mu$m$^{-1}$, logarithmic colour scale), together with streamlines, and (d) the electrochemical potential $\mu^\star/e$ with its equipotential lines. From the same solutions the four-terminal resistances $R_{xy}=V_{14}/I$ and $R_{xx}=V_{12}/I$, the interior-contact voltage $R_{AB}=V_{AB}/I$ and the voltage between an interior and a perimeter contact, $R_{B2}=V_{B2}/I$, are extracted for the S--D excitation, and the transfer coefficients $V_{AB}/I_{AB}$, $V_{B2}/I_{AB}$, $V_{23}/I_{AB}$ and $V_{12}/I_{AB}$ for the A--B excitation; their dependence on $B$ across the plateau is collected in Figs.~\ref{fig:sweep} and \ref{fig:sweepab}.

\section{Results}
\label{sec:results}

\subsection{Electron density at zero magnetic field}
\label{sec:density}

\begin{figure}
\includegraphics[width=0.9\textwidth]{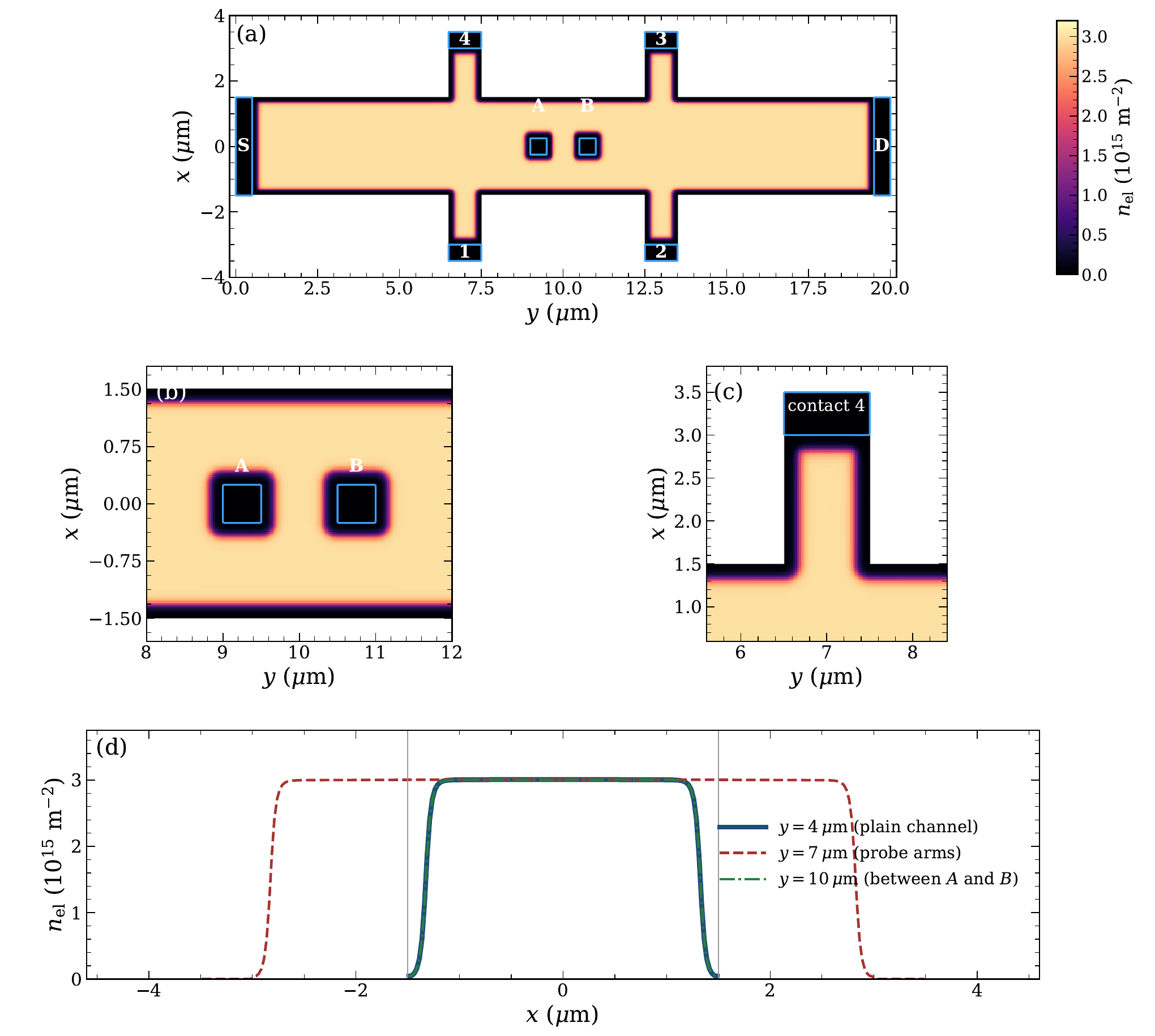}
\caption{Self-consistent electron density at $B=0$. (a) Full device; the bulk density is $n_0=3.0\times10^{15}\,{\rm m}^{-2}$. (b) Zoom onto the interior contacts A and B and their depletion halos. (c) Zoom onto the junction between the channel and the arm of probe 4; the arm is fully populated. (d) Transverse density profiles $n_{\rm el}(x)$ at $y=4\,\mu$m (plain channel), $y=7\,\mu$m (through the probe arms) and $y=10\,\mu$m (between A and B). The vertical lines mark the lithographic edges $x=\pm1.5\,\mu$m. The profiles at $y=4$ and $10\,\mu$m coincide.}
\label{fig:density}
\end{figure}

Figure~\ref{fig:density}(a) shows the self-consistent electron density at $B=0$. In the plain channel the density is flat at $n_0=3.0\times10^{15}\,{\rm m}^{-2}$ and drops to zero over a depletion length of roughly $0.3\,\mu$m at the mesa edges, so that the electronic width of the channel is about $2.4\,\mu$m (dashed lines in Fig.~\ref{fig:geometry}). The etched edges produce a steep profile, as expected for mesa-defined samples whose side surfaces pin the Fermi level~\cite{Arslan08,Gerhardts08}. The probe arms are fully populated up to their contacts [Fig.~\ref{fig:density}(c)], i.e. the arms are wide enough that no pinch-off occurs at their junction with the channel; this is essential for the perimeter probes to act as proper voltage terminals.

The interior contacts are surrounded by depletion halos of similar width [Fig.~\ref{fig:density}(b)], the ``electron-poor region'' in front of a metallic contact~\cite{Eksi10,Dahlem10}. Because A and B are $1\,\mu$m apart and $0.95\,\mu$m from the electronic edge, their depletion regions do not overlap with each other or with the edge depletion. This is demonstrated quantitatively in Fig.~\ref{fig:density}(d): the transverse profiles at $y=4\,\mu$m (plain channel) and at $y=10\,\mu$m (between A and B) are indistinguishable, while the profile at $y=7\,\mu$m through the probe arms extends to $|x|\approx 2.8\,\mu$m. The interior contacts therefore modify the density only locally, and the channel between them remains a bulk-like 2DES; the difference from the $10\,\mu$m bar is one of degree, in that here the contact-induced depletion occupies roughly one third of the channel width at the position of A and B.

\subsection{Filling-factor landscape at $\nu_{\rm nom}=2.1$}
\label{sec:filling}

\begin{figure}
\includegraphics[width=0.9\textwidth]{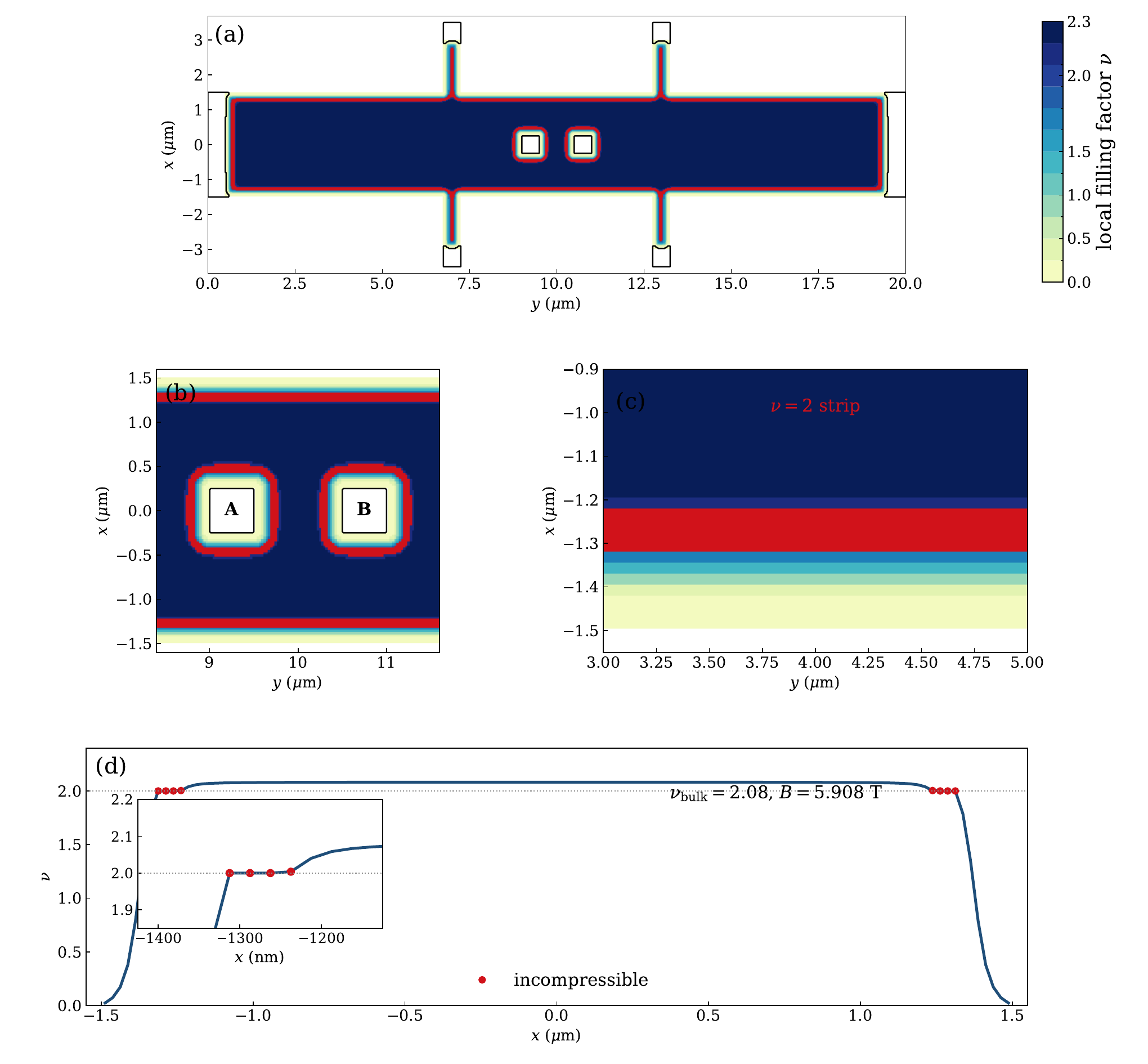}
\caption{Local filling factor at $\nu_{\rm nom}=2.1$ ($B=\SI{5.908}{T}$, $\nu_{\rm bulk}=2.08$, $T=\SI{1.5}{K}$); incompressible cells in red. (a) Full device. (b) Zoom onto the interior contacts, each enclosed by a closed $\nu=2$ ring. (c) Zoom onto the lower edge of the plain channel between $y=3$ and $5\,\mu$m. (d) Transverse profile $\nu(x)$ at $y=4\,\mu$m; red points: incompressible cells. Inset: the $\nu=2$ plateau at the lower edge on a nanometre scale.}
\label{fig:filling}
\end{figure}

At a magnetic field slightly below that of the plateau centre, $\nu_{\rm nom}=2.1$ ($B=\SI{5.908}{T}$, $\nu_{\rm bulk}=2.08$), the local filling factor develops the structure shown in Fig.~\ref{fig:filling}. The bulk of the channel and the probe arms are compressible at $\nu\approx2.08$. Along all boundaries of the 2DES---the channel edges, the arms and the perimeter of each interior contact---the filling factor passes through the value $2$, and the self-consistent screening produces a narrow incompressible strip wherever this happens~\cite{Chklovskii92,Lier94} [red in Figs.~\ref{fig:filling}(a)--(c)].

Figure~\ref{fig:filling}(c) magnifies the lower edge of the plain channel. The $\nu=2$ strip is centred at $x\approx-1.27\,\mu$m, about $\SI{230}{nm}$ inside the lithographic edge, and is four cells ($\SI{100}{nm}$) wide; on the $10\,$nm grid it is $\SI{80}{nm}$ wide (Sec.~\ref{sec:robust}). The transverse profile $\nu(x)$ in Fig.~\ref{fig:filling}(d), and its inset, show the plateau at $\nu=2$ that results from the enhanced screening in the compressible regions on either side and the pinning of the density in the incompressible region. The strip is several magnetic lengths ($\ell_B\approx\SI{10.5}{nm}$) and about two Fermi wavelengths wide, so that it survives the coarse-graining of Sec.~\ref{sec:tfpa} and can be treated by the local conductivity model; strips narrower than $\lambda_F$ would be ``evanescent''~\cite{Sailer10,Kendirlik13} and transparent to scattering. The points marked ``incompressible'' in Fig.~\ref{fig:filling}(d) are those where the criterion of Sec.~\ref{sec:tfpa} is satisfied.

The crucial feature for the interior contacts is visible in Fig.~\ref{fig:filling}(b): each of A and B is surrounded by a \emph{closed} $\nu=2$ ring, separated from the edge strips by compressible bulk. Topologically, the sample at this field consists of a compressible bulk that is bounded on the outside by the edge strips and that contains two compressible islands (the depletion halos of A and B) bounded by the rings; the incompressible area of the sample consists of exactly three connected regions, the edge strip ($4.7\,\mu$m$^2$, continuous around the whole perimeter including the arms) and the two rings ($0.3\,\mu$m$^2$ each). This is the configuration sketched as panel (c) in Fig.~2 of Ref.~\cite{Kendirlik17} and obtained for the wider bar in Fig.~1(a,b) of Ref.~\cite{Yildiz14}. In the present model the rings have the same width as the edge strips within the resolution of the grid, because the same donor-free rim $l_{\rm pin}$ is used in front of a contact window and at the etched edge; a smoother contact halo would make them narrower~\cite{Yildiz14}.

\subsection{Current distribution across the $\nu=2$ plateau}
\label{sec:current}

Figures~\ref{fig:cur18}--\ref{fig:cur22} present the filling factor, the conductivity, the current and the electrochemical potential for five fields, in order of decreasing $B$, corresponding to $\nu_{\rm nom}=1.8$, $1.9$, $2.0$, $2.1$ and $2.2$, i.e. $B=\SI{6.89}{}$, $\SI{6.53}{}$, $\SI{6.20}{}$, $\SI{5.91}{}$ and $\SI{5.64}{T}$ ($\nu_{\rm bulk}=1.82$, $1.92$, $2.00$, $2.08$ and $2.19$). The current $I=\SI{1}{nA}$ is imposed between S and D. The resistances quoted below are those of the same solutions and are plotted against $B$ in Fig.~\ref{fig:sweep}; the two-terminal resistance $R_{AB}^{(AB)}=V_{AB}/I_{AB}$ for the A--B excitation is taken from Fig.~\ref{fig:sweepab}. Inside the plateau both $R_{xx}$ and $R_{AB}$ for the S--D excitation are determined by transport across incompressible regions and are therefore $\epsilon$-limited (Sec.~\ref{sec:ohm}); for these we quote only bounds.

\begin{figure}
\includegraphics[width=0.9\textwidth]{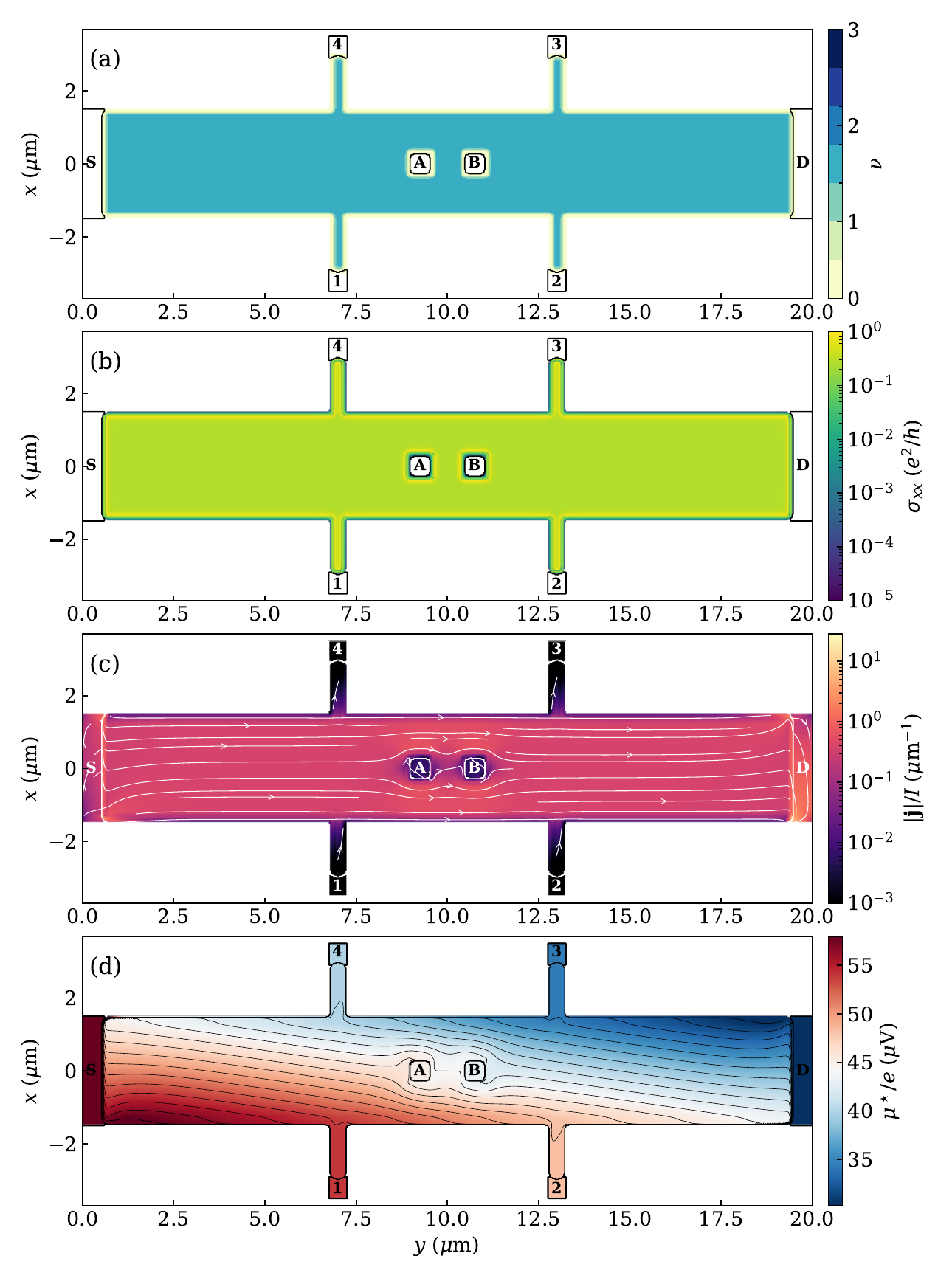}
\caption{$\nu_{\rm nom}=1.8$ ($B=\SI{6.89}{T}$, $\nu_{\rm bulk}=1.82$, $I=\SI{1}{nA}$ between S and D, $T=\SI{1.5}{K}$). (a) Local filling factor $\nu(x,y)$; no $\nu=2$ region is present. (b) $\sigma_{xx}$ (logarithmic colour scale). (c) Normalized current density $|\bm j|/I$ (colour, logarithmic scale) with streamlines. (d) Electrochemical potential $\mu^\star/e$ with equipotential lines.}
\label{fig:cur18}
\end{figure}

\subsubsection{$\nu_{\rm nom}
=1.8$ (Fig.~\ref{fig:cur18}).} At the highest field, above the high-field end of the plateau, the filling factor is $1.82$ in the bulk and below $2$ everywhere, so that no incompressible region exists [Fig.~\ref{fig:cur18}(a)]; $\sigma_{xx}\approx0.5\,e^2/h$ is uniform over the 2DES [Fig.~\ref{fig:cur18}(b)]. The sample is a classical Hall conductor: the current spreads over the entire channel with $|j_y|/I\approx0.38\,\mu$m$^{-1}$ across the electronic width of $2.4\,\mu$m, the profile following the density profile (Drude-like, $j_y\propto n_{\rm el}$) [Fig.~\ref{fig:cur18}(c)]. The equipotentials of $\mu^\star/e$ are tilted straight lines that cross the channel [Fig.~\ref{fig:cur18}(d)]; the Hall voltage of $\SI{14}{\mu V}$ across the channel and the longitudinal drop of $\SI{5.6}{\mu V}$ between the probes give $R_{xy}$ 9\% above $h/2e^2$ and $R_{xx}=\SI{5.6}{k\ohm}$. The streamlines are deflected around the depletion halos of A and B and the equipotentials bulge around them, so that both contacts sit at potentials that differ from the local Hall potential of the undisturbed channel; $R_{AB}=\SI{1.7}{k\ohm}$, about $R_{xx}/3$, and $R_{AB}^{(AB)}=1.6\times10^{5}\,\Omega$.

\begin{figure}
\includegraphics[width=0.9\textwidth]{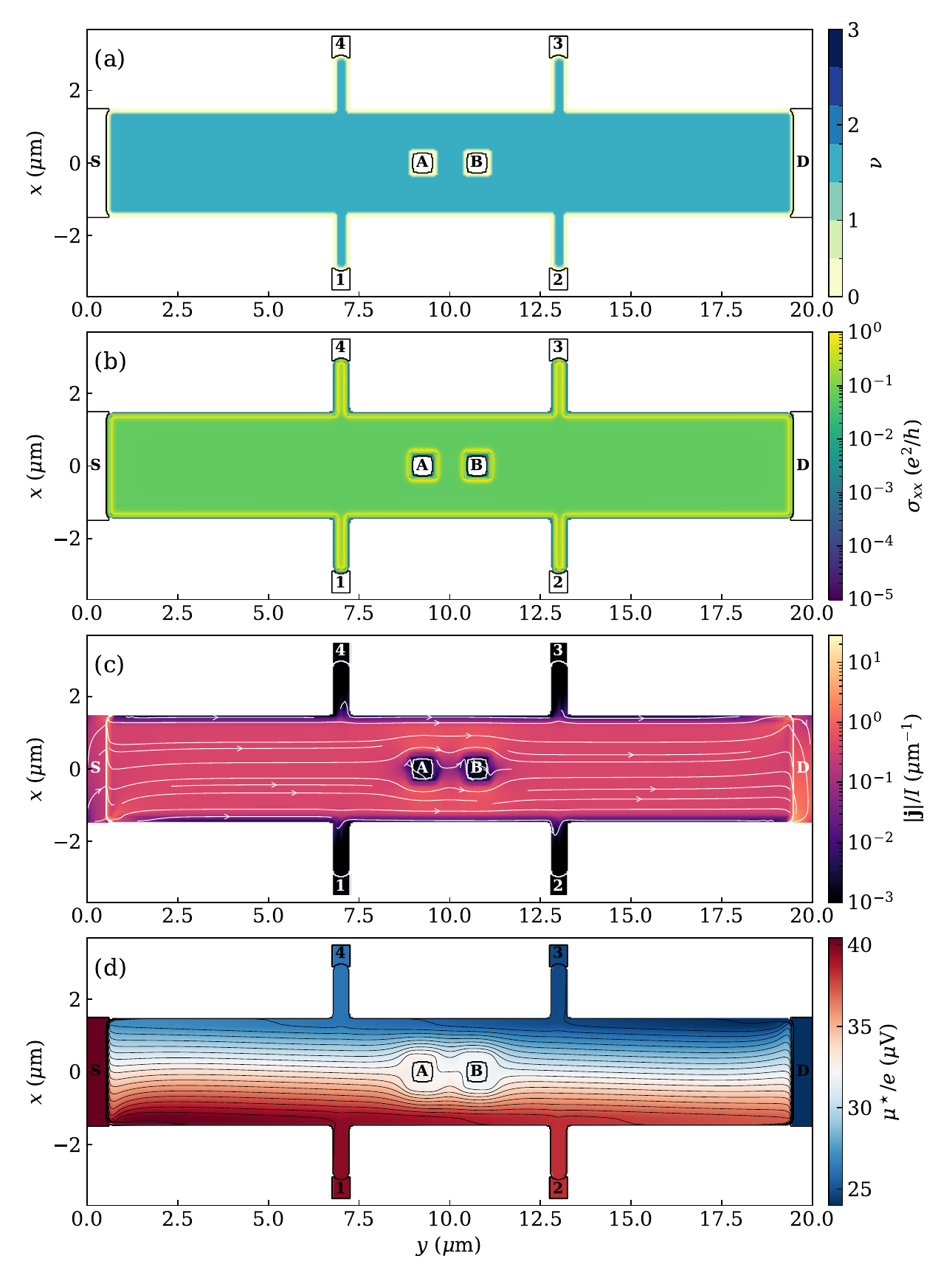}
\caption{$\nu_{\rm nom}=1.9$ ($B=\SI{6.53}{T}$, $\nu_{\rm bulk}=1.92$), high-field side of the plateau. Panels as in Fig.~\ref{fig:cur18}. No $\nu=2$ regions are present; the current spreads over the whole compressible channel and is deflected around A and B; the Hall potential varies linearly across the channel.}
\label{fig:cur19}
\end{figure}

\subsubsection{$\nu_{\rm nom}
=1.9$ (Fig.~\ref{fig:cur19}).} On the high-field side of the plateau the bulk filling factor is $1.92$, still below $2$ everywhere in the interior, and no $\nu=2$ regions are formed [Fig.~\ref{fig:cur19}(a,b)]. The whole sample is compressible and dissipative, and the current distributes over the channel as in a classical Hall conductor~\cite{Guven03,Eksi10}: $|j_y|/I\approx0.40\,\mu$m$^{-1}$ across the electronic width. In the interior region the streamlines in Fig.~\ref{fig:cur19}(c) are deflected around the depletion halos of A and B and rejoin behind them, with the current density enhanced in the two lanes between the contacts and the edges; this is the behaviour found by Gro{\ss} and Gerhardts for antidots near the breakdown of the IQHE~\cite{Gross98}. The Hall potential in Fig.~\ref{fig:cur19}(d) varies linearly across the channel, the ``type I'' profile of the scanning-probe experiments~\cite{Ahlswede01,Siddiki04}, and is only locally distorted around the contacts; $R_{xx}=\SI{1.3}{k\ohm}$ and $R_{xy}$ is 4\% above $h/2e^2$. The contacts are thus embedded in a resistive medium and are galvanically connected to the current path; in the language of Ref.~\cite{Kendirlik17} the bulk is a poorly conducting metal modelled by the series resistances $R_C$ and $R_B$ (their panel f), and a voltage $V_{AB}$ of the order of $\rho_\ell I\,(y_B-y_A)/W$ is expected between the contacts. Here $R_{AB}=\SI{450}{\ohm}$ and $R_{AB}^{(AB)}=2.2\times10^{5}\,\Omega$.

\begin{figure}
\includegraphics[width=0.9\textwidth]{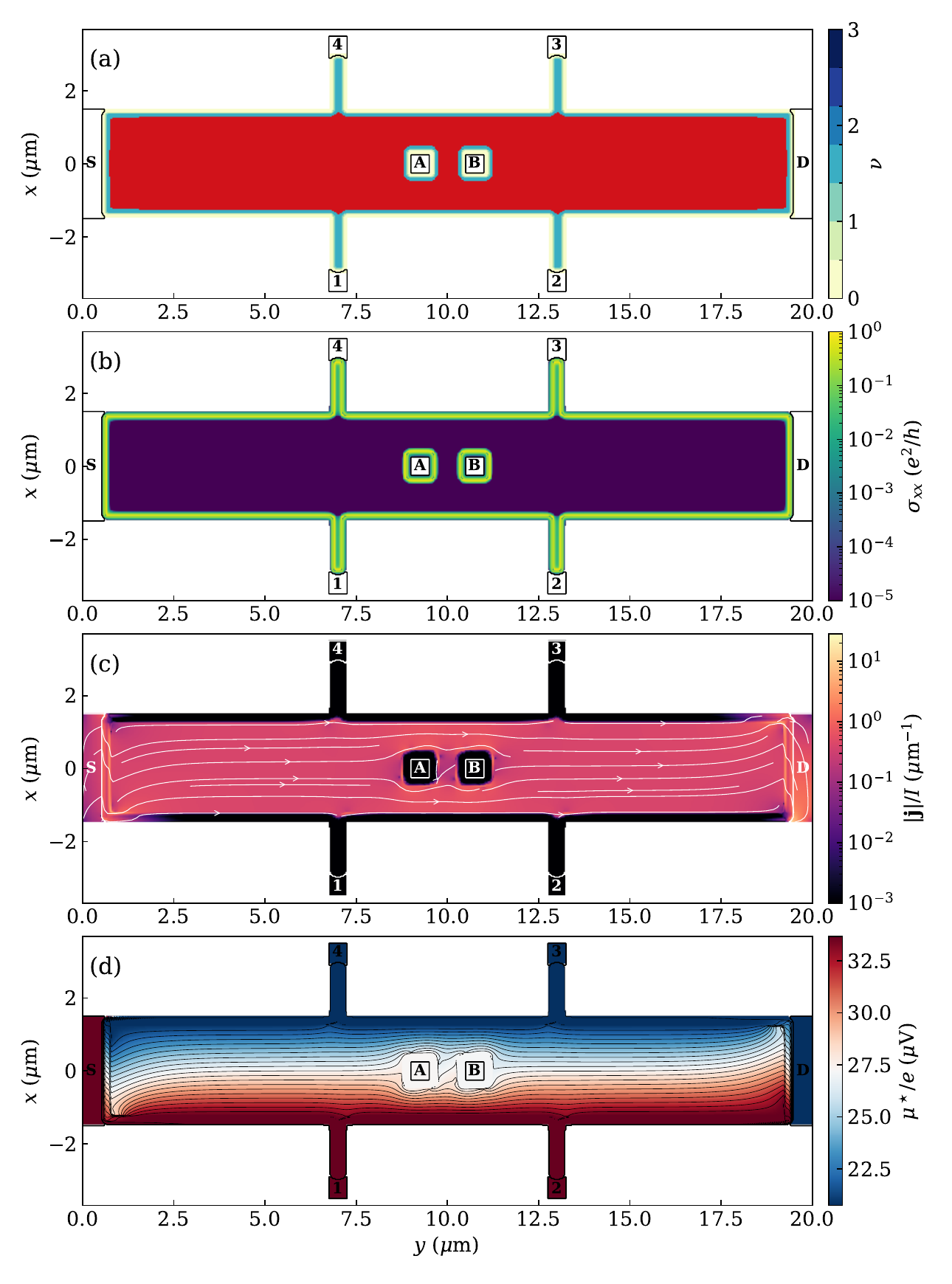}
\caption{$\nu_{\rm nom}=2.0$ ($B=\SI{6.20}{T}$, $\nu_{\rm bulk}=2.00$), plateau centre. Panels as in Fig.~\ref{fig:cur18}. The whole channel is incompressible, bounded by narrow compressible rims at the edges and around A and B; the current is carried by the incompressible bulk with a nearly uniform $|\bm j|/I$; the Hall potential drops across the whole channel.}
\label{fig:cur20}
\end{figure}

\subsubsection{$\nu_{\rm nom}
=2.0$ (Fig.~\ref{fig:cur20}).} At the plateau centre the entire channel, the probe arms and the region between and around the interior contacts have $\nu=2$: the sample is a single incompressible region of $45\,\mu$m$^2$ bounded by narrow compressible rims at the edges and around A and B [Fig.~\ref{fig:cur20}(a)], and $\sigma_{xx}$ vanishes everywhere except in these rims [Fig.~\ref{fig:cur20}(b)]. This is the incompressible-bulk configuration of the single-particle theories, which in the screening theory occurs only in a subinterval of the plateau~\cite{Siddiki04,Yildiz14}; a configuration of this kind was also reported for a wider bar in Ref.~\cite{Sirt23}. The current is now carried by the incompressible bulk as a dissipationless Hall current: $|\bm j|/I$ is nearly uniform across the channel [Fig.~\ref{fig:cur20}(c)], and the corresponding Hall potential in Fig.~\ref{fig:cur20}(d) drops across the whole width of the sample by $\SI{13}{\mu V}$, the ``type II'' profile of the scanning-probe experiments~\cite{Ahlswede01,Siddiki04}; $R_{xy}=h/2e^2$ to $10^{-6}$ for every value of $\epsilon$ and grid tested, and $R_{xx}\propto\epsilon$ ($\SI{0.2}{\ohm}$ for $\epsilon=10^{-5}$). Because $\sigma_\ell=0$ in the bulk, the current is not steered by the density profile but by the requirement $\nabla\cdot\bm j=0$ with $\bm j\perp\bm E$; the streamlines bend smoothly around the interior contacts and around the junctions with the arms. The compressible rims around A and B are the only dissipative regions in the vicinity of the contacts and carry almost no current. The interior contacts are equipotential islands floating in the Hall potential of an insulating bulk. Their potential difference $V_{AB}$ for the S--D current is set by the residual conductance $\epsilon e^2/h$ of the bulk and by the discretized geometry of the two rims: it changes from $\SI{20}{\ohm}$ to $\SI{390}{\ohm}$ between $\epsilon=10^{-3}$ and $10^{-6}$ and changes sign on the $10\,$nm grid, so that only the bound $|R_{AB}|\lesssim10^{-2}R_{xy}$ is meaningful. For a current injected at A and extracted at B, on the other hand, the contacts are separated by the incompressible bulk, the quantum capacitance $C=e^2D_T(E_F)A$ vanishes for an ideal sample at $T\to0$, and the impedance diverges~\cite{Yildiz14,Kendirlik17}; in the model $R_{AB}^{(AB)}\propto1/\epsilon$ ($6\times10^{8}\,\Omega$ for $\epsilon=10^{-5}$), three orders of magnitude above its value in the compressible regime. This is the origin of the saturated $V_{AB}=V_{AB}^{\max}$ at the high-field end of the saturation interval in the experiment [panel (d) of Fig.~2 in Ref.~\cite{Kendirlik17}].

\begin{figure}
\includegraphics[width=0.9\textwidth]{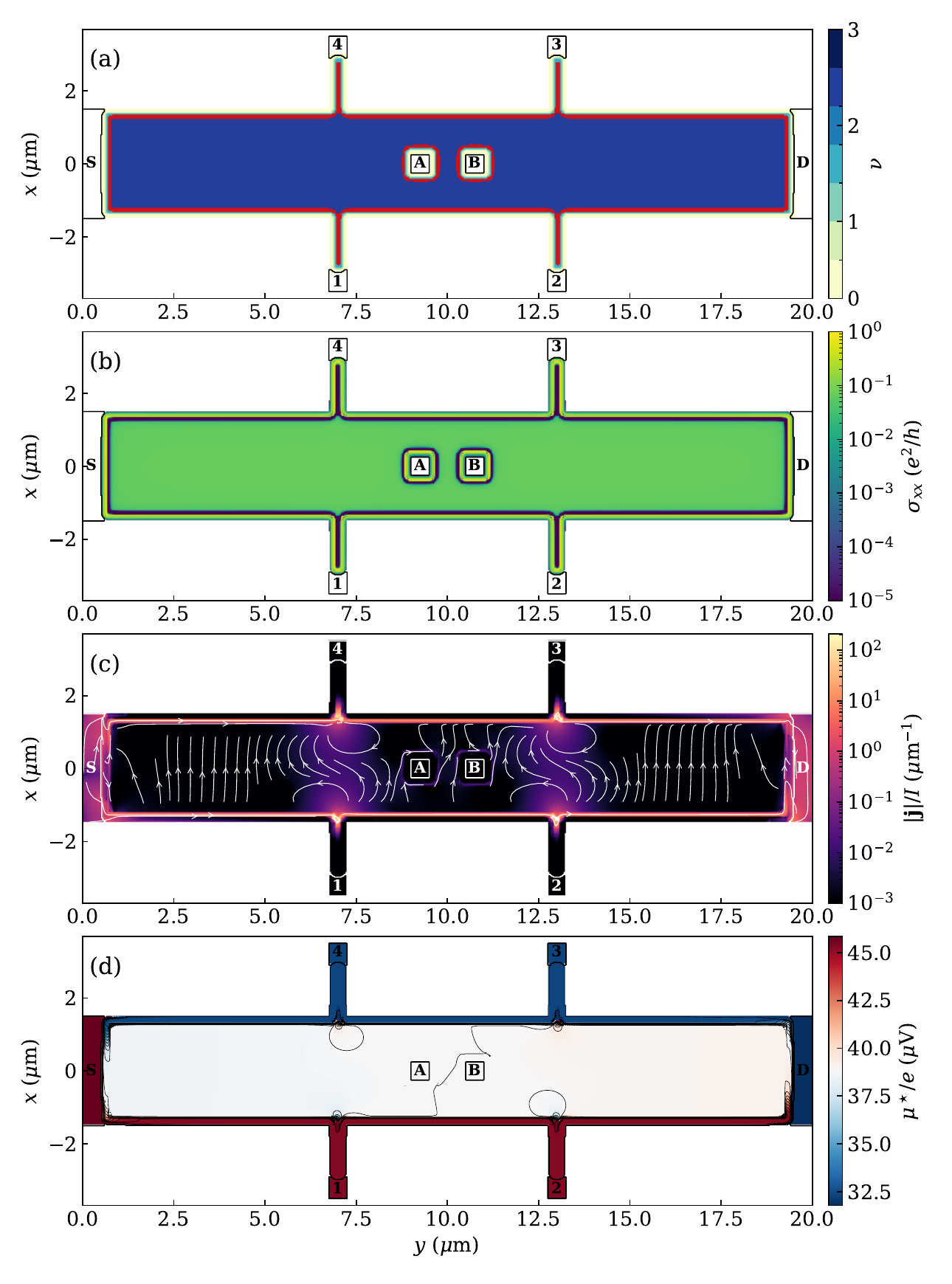}
\caption{$\nu_{\rm nom}=2.1$ ($B=\SI{5.91}{T}$, $\nu_{\rm bulk}=2.08$). Panels as in Fig.~\ref{fig:cur18}. The bulk is compressible and $\nu=2$ strips line the edges, the probe arms and the perimeters of A and B, where they form closed rings; the current flows in the two edge strips; the Hall potential drops across the edge strips and is flat in the bulk, including at A and B.}
\label{fig:cur21}
\end{figure}

\subsubsection{$\nu_{\rm nom}
=2.1$ (Fig.~\ref{fig:cur21}).} Slightly above $\nu=2$ the bulk becomes compressible ($\nu_{\rm bulk}=2.08$) while the boundaries retain their $\nu=2$ strips, as discussed in Sec.~\ref{sec:filling} [Figs.~\ref{fig:filling} and \ref{fig:cur21}(a,b)]. The current distribution changes qualitatively: the compressible bulk is short-circuited by the two dissipationless edge strips, and the current is carried by them [bright lines along the boundary in Fig.~\ref{fig:cur21}(c), where $|\bm j|/I$ reaches $10\,\mu$m$^{-1}$]. Integrating $j_y$ across the channel at $y=6\,\mu$m, $99.9\%$ of $I$ flows in the $\nu=2$ cells and their immediate flanks and $0.08\%$ in the compressible bulk; the streamlines that remain visible in the bulk are those of this residual current, which closes in loops between the strips and carries no net current. The edge strips follow the entire boundary of the 2DES, including the arms of the probes, and are continuous from S to D; the Hall potential drops entirely across them and is flat elsewhere [``type III'' profile~\cite{Ahlswede01,Siddiki04}, Fig.~\ref{fig:cur21}(d)], and $R_{xy}=h/2e^2$ to $10^{-4}$ while $R_{xx}$ is $\epsilon$-limited ($\SI{0.7}{\ohm}$ on the $25\,$nm grid, $\SI{15}{\ohm}$ on the $10\,$nm grid, Sec.~\ref{sec:robust}). The closed rings around A and B are disconnected from the edge strips and enclose no source of current; they carry no net current [Fig.~\ref{fig:cur21}(c)] and act as insulating barriers that separate the compressible islands around the contacts from the compressible bulk. This is the configuration in which no non-equilibrium current flows in the strips encircling the inner contacts, as anticipated in the sketch of Ref.~\cite{Kendirlik17} and in the calculation of Ref.~\cite{Sirt23}. The interior contacts are thus electrically isolated from both the current-carrying edges and, through the rings, from the bulk. For an S--D current $V_{AB}$ is again an $\epsilon$-limited residual, $|R_{AB}|<\SI{30}{\ohm}$, i.e. the contacts lie in the flat part of the potential landscape of Fig.~\ref{fig:cur21}(d); for an A--B current the rings act as the capacitors $C_{IS}$ of Ref.~\cite{Kendirlik17} [their panel (c)], and $R_{AB}^{(AB)}$ is $\epsilon$-limited at $9\times10^{7}\,\Omega$, so that $V_{AB}$ stays saturated even though the bulk between the contacts is compressible. The Hall voltage at the perimeter remains quantized because the edge strips decouple the Hall probes~\cite{Siddiki04,Kendirlik17}.

\begin{figure}
\includegraphics[width=0.9\textwidth]{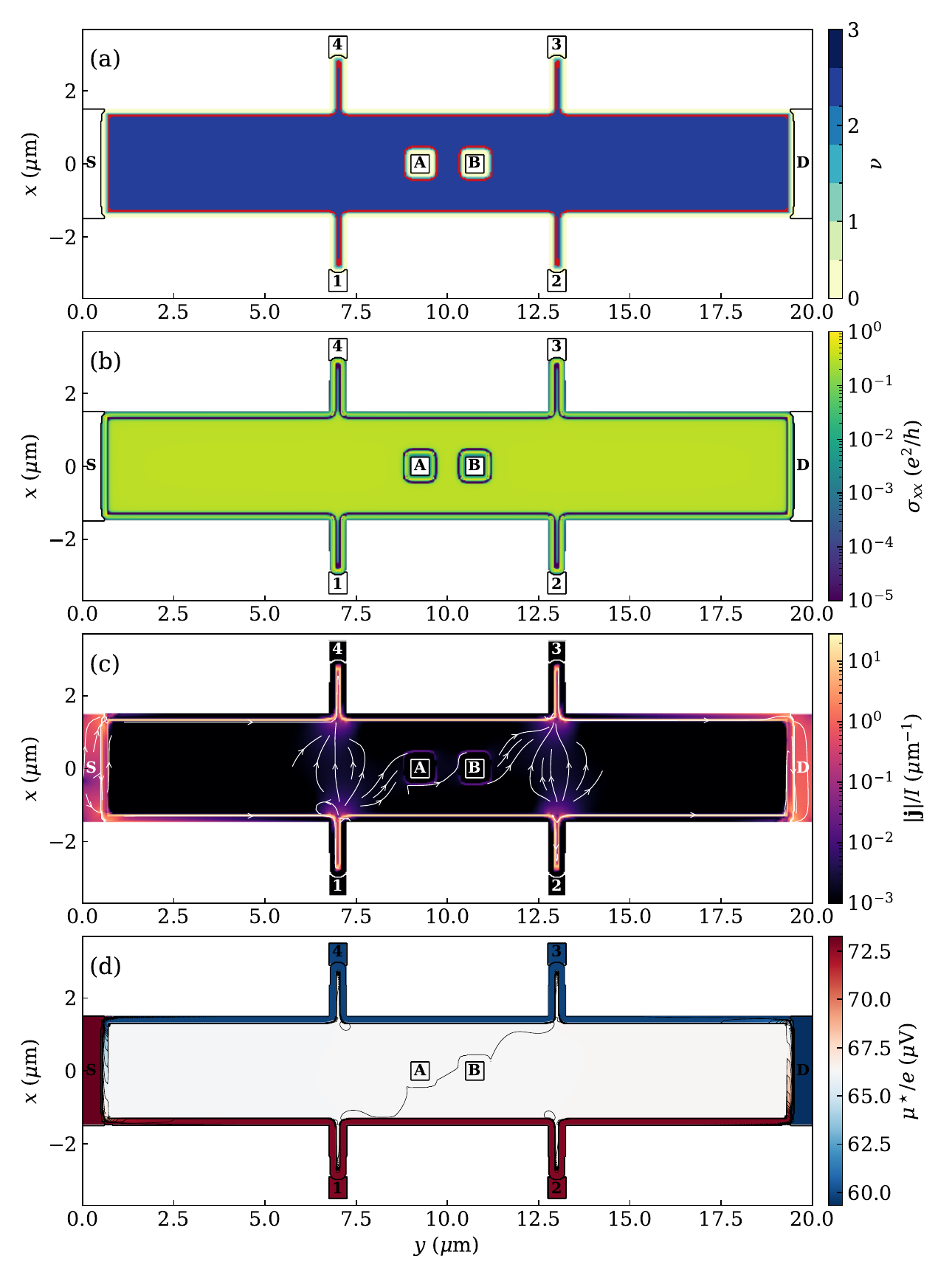}
\caption{$\nu_{\rm nom}=2.2$ ($B=\SI{5.64}{T}$, $\nu_{\rm bulk}=2.19$), low-field end of the plateau. Panels as in Fig.~\ref{fig:cur18}. The edge strips and the rings around A and B have moved outward and become narrower; the current remains in the edge strips, with a weak leakage into the bulk at the concave corners of the probe-arm junctions; the Hall potential is flat in the bulk and around the interior contacts.}
\label{fig:cur22}
\end{figure}

\subsubsection{$\nu_{\rm nom}
=2.2$ (Fig.~\ref{fig:cur22}).} Further towards the low-field end of the plateau ($\nu_{\rm bulk}=2.19$) the edge strips move outward and become narrower, and the rings around A and B shrink towards the contacts [Fig.~\ref{fig:cur22}(a,b)], in accordance with the $B$ dependence of the strip width in the CSG and TFPA descriptions~\cite{Chklovskii92,Siddiki04}: the incompressible area is $2.7\,\mu$m$^2$ for the edge strip and $0.16\,\mu$m$^2$ per ring, about half of the values at $\nu_{\rm nom}=2.1$, and the rings are two cells ($\SI{50}{nm}$, $\approx\lambda_F$) wide. The current is still carried by the edge strips, but a leakage into the compressible bulk is now visible at the concave corners of the probe-arm junctions, where the strips have to bend [Fig.~\ref{fig:cur22}(c)]: at $y=6\,\mu$m, $0.19\%$ of $I$ flows in the bulk, against $0.08\%$ at $\nu_{\rm nom}=2.1$, and $R_{xx}$ rises to $\SI{12}{\ohm}$ ($\SI{79}{\ohm}$ on the $10\,$nm grid) while $R_{xy}=h/2e^2$ to $3\times10^{-4}$. This is the precursor of the breakdown of the plateau on its low-field side: as $B$ is decreased further the strips become narrower than $\lambda_F$, the local conductivity model no longer supports a dissipationless channel, and the current returns to the bulk~\cite{Siddiki04}. The interior contacts remain isolated at $\nu_{\rm nom}=2.2$ [flat potential around them in Fig.~\ref{fig:cur22}(d), $|R_{AB}|<\SI{15}{\ohm}$, $R_{AB}^{(AB)}=1.4\times10^{7}\,\Omega$], but the thinning of their rings, which are the narrowest incompressible regions of the sample, indicates that they become leaky before the edge strips do; since the rings are only about $\lambda_F$ wide here, their insulating character is an upper bound of the model. In the experiment this is precisely the sequence observed at the low-field end of the saturation interval [panel (e$_1$) of Fig.~2 in Ref.~\cite{Kendirlik17}]: $V_{AB}$ drops below $V_{AB}^{\max}$ while $V_H$ is still quantized, because the encircling strips become thermodynamically compressible earlier than the edge strips. The isolation of the interior contacts is therefore expected to end at a higher field than the plateau itself, and to be more sensitive to temperature~\cite{Kendirlik17}; both expectations are confirmed by the sweeps and the sensitivity tests of Sec.~\ref{sec:discussion}.

\section{Discussion}
\label{sec:discussion}

\subsection{Three transport regimes and the interior contacts}

The five calculations map out the three regimes of the screening theory within a narrow interval of the magnetic field. Table~\ref{tab:regimes} summarizes them together with the resulting state of the interior contacts and the corresponding sketch of Ref.~\cite{Kendirlik17}.

\begin{table}
\caption{Transport regimes around the $\nu=2$ plateau in the order of decreasing $B$: state of the bulk, region carrying the source--drain current, coupling of the interior contacts A and B to that current, the isolation ratio $\mathcal{R}=R_{AB}^{(AB)}(\nu_{\rm nom})/R_{AB}^{(AB)}(1.8)$ for $\epsilon=10^{-5}$ (values in the isolated and floating regimes scale as $1/\epsilon$), and the corresponding panel of Fig.~2 in Ref.~\cite{Kendirlik17}.}
\label{tab:regimes}
\begin{indented}
\item[]\begin{tabular}{@{}lllccc}
\br
$\nu_{\rm nom}$ & bulk & current & A, B & $\mathcal{R}$ & Ref.~\cite{Kendirlik17} \\
\mr
1.8 & compr. & bulk & connected & 1 & (f) \\
1.9 & compr. & bulk & connected & 1.4 & (f) \\
2.0 & incompr. & bulk (Hall) & floating & $4\times10^{3}$ & (d) \\
2.1 & compr. & edge strips & isolated & $6\times10^{2}$ & (c) \\
2.2 & compr. & edge strips & isolated (rings thin) & $9\times10^{1}$ & (e$_1$) \\
\br
\end{tabular}
\end{indented}
\end{table}

\begin{figure}
\includegraphics[width=0.9\textwidth]{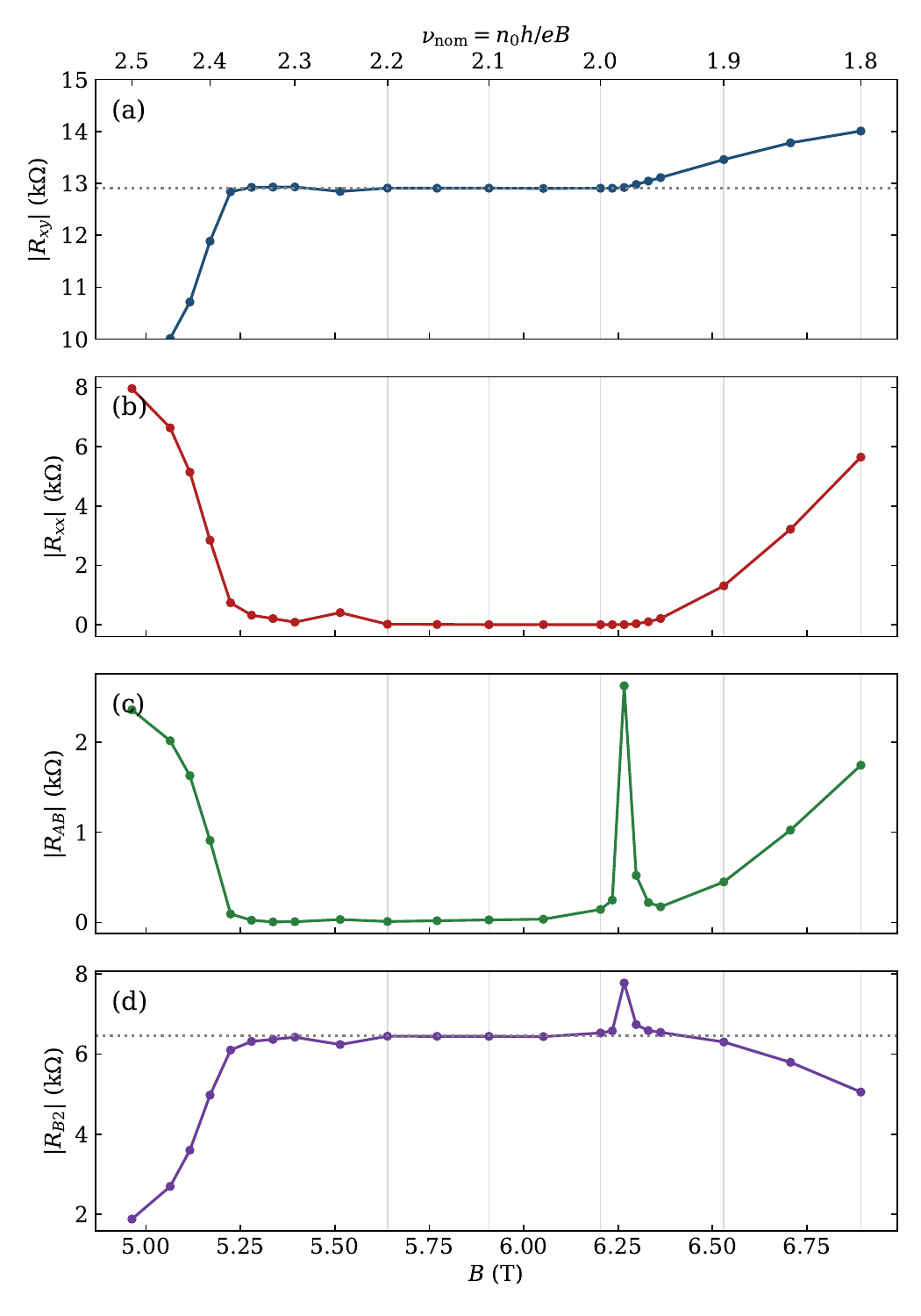}
\caption{Magnetic-field dependence of the four-terminal resistances for a dc source--drain current $I=\SI{1}{nA}$ at $T=\SI{1.5}{K}$ across the $\nu=2$ plateau (magnitudes); the top axis gives $\nu_{\rm nom}=n_0h/eB$ and the vertical lines mark the five fields of Figs.~\ref{fig:cur18}--\ref{fig:cur22}. (a) $|R_{xy}|$ (dotted line: $h/2e^2$); (b) $|R_{xx}|$; (c) $|R_{AB}|$, $R_{AB}=V_{AB}/I$; (d) $|R_{B2}|$, $R_{B2}=V_{B2}/I$, between interior contact B and perimeter probe 2 (dotted line: $h/4e^2$).}
\label{fig:sweep}
\end{figure}

\begin{figure}
\includegraphics[width=0.9\textwidth]{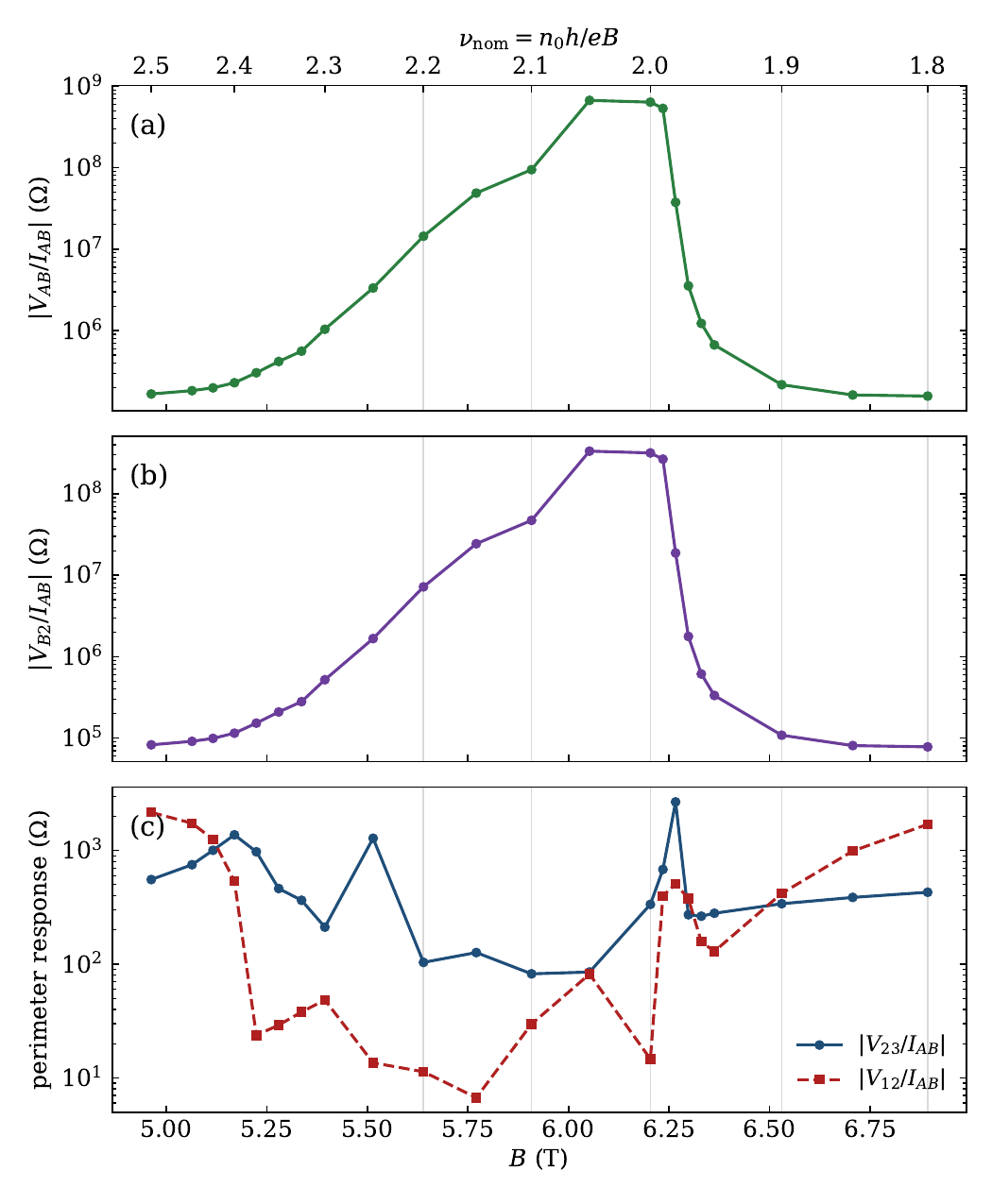}
\caption{Transfer coefficients for a dc current $I_{AB}=\SI{1}{nA}$ injected at A and extracted at B, all other contacts floating ($T=\SI{1.5}{K}$, $\epsilon=10^{-5}$). (a) Two-terminal resistance $|V_{AB}/I_{AB}|$; (b) $|V_{B2}/I_{AB}|$; (c) perimeter response $|V_{23}/I_{AB}|$ (Hall probes) and $|V_{12}/I_{AB}|$ (longitudinal probes). The values of (a) and (b) inside the plateau scale as $1/\epsilon$.}
\label{fig:sweepab}
\end{figure}

The magnetic-field sweeps in Figs.~\ref{fig:sweep} and \ref{fig:sweepab} place the five snapshots in context. They were computed on the same grid and at the same temperature as the maps, in steps of $\Delta\nu_{\rm nom}=0.05$ refined to $0.01$ at the high-field end of the plateau and to $0.025$ at its low-field end, each field starting from its neighbour, and for the two excitations separately.

For the S--D excitation (Fig.~\ref{fig:sweep}) the perimeter resistances show the expected plateau: $R_{xy}$ is quantized at $h/2e^2$ within $0.5\%$ between $\SI{5.2}{}$ and $\SI{6.27}{T}$, i.e. for $\nu_{\rm nom}$ between $2.375$ and $1.98$, and $R_{xx}$ is below $\SI{15}{\ohm}$ between $\SI{5.64}{}$ and $\SI{6.24}{T}$; towards lower fields $R_{xx}$ rises in steps as the strips narrow ($\SI{80}{}$--$\SI{740}{\ohm}$ for $\nu_{\rm nom}=2.3$--$2.375$) before the plateau breaks down between $\nu_{\rm nom}=2.375$ and $2.40$, where the edge strips fragment into more than thirty disconnected pieces, and towards higher fields it rises smoothly to $\SI{5.6}{k\ohm}$ at $\nu_{\rm nom}=1.8$ together with the classical increase of $R_{xy}$. The plateau is asymmetric about $\nu_{\rm nom}=2$, extending mainly to its low-field side, as in the screening theory of the wide bar~\cite{Siddiki04}. The interior-contact voltage behaves differently from the perimeter voltages: $|R_{AB}|$ is below $\SI{40}{\ohm}$ throughout the edge-strip regime and of order $10^{2}\,\Omega$ (both $\epsilon$-limited) in the incompressible-bulk regime, and grows monotonically on the high-field side, where it follows $R_{xx}$ with a ratio $R_{AB}/R_{xx}\approx1/3$ set by the geometry ($y_B-y_A$ relative to the probe separation and the deflection of the current around the contacts). $|R_{B2}|$, the voltage between an interior contact and a perimeter probe, approaches one half of the quantized Hall resistance, $h/4e^2$, over the whole plateau [dotted line in Fig.~\ref{fig:sweep}(d); within 3\% between $\SI{5.3}{}$ and $\SI{6.4}{T}$]. This is a consequence of the symmetric geometry: the interior contact on the axis of the channel samples the midpoint of the Hall potential drop whenever the bulk is either incompressible or isolated from the contact by its ring, and it departs from $h/4e^2$ on both sides of the plateau together with $R_{xy}$. A calculation with the contacts displaced by $0.3\,\mu$m from the axis (Sec.~\ref{sec:robust}) confirms this reading: in the incompressible-bulk regime $R_{B2}/R_{xy}$ moves from $0.50$ to $0.67$, whereas in the edge-strip regime, where the potential is flat across the bulk, it stays at $0.50$.

For the A--B excitation (Fig.~\ref{fig:sweepab}) the transfer coefficients reveal the isolation of the contacts directly. The two-terminal resistance $R_{AB}^{(AB)}=V_{AB}/I_{AB}$ is $1.6$--$2.3\times10^{5}\,\Omega$ outside the plateau on both sides, where the injected current spreads through the compressible bulk and the value is independent of $\epsilon$ (Sec.~\ref{sec:robust}). Inside the plateau it rises by orders of magnitude: gradually on the low-field side, from $4\times10^{5}\,\Omega$ at $\nu_{\rm nom}=2.35$ through $10^{6}$ at $2.30$, $1.4\times10^{7}$ at $2.2$ and $9\times10^{7}\,\Omega$ at $2.1$ to $6$--$7\times10^{8}\,\Omega$ at $2.05$--$1.99$, because the rings, which carry the whole two-terminal resistance in the edge-strip regime, widen continuously; and it falls abruptly on the high-field side, from $5\times10^{8}\,\Omega$ at $\nu_{\rm nom}=1.99$ to $4\times10^{7}$ at $1.98$, $3.5\times10^{6}$ at $1.97$ and $7\times10^{5}\,\Omega$ at $1.95$, because the incompressible bulk disappears within $\Delta B/B\approx0.5\%$ (the number of incompressible regions drops from one to zero between $1.98$ and $1.97$). The same behaviour is seen in $V_{B2}/I_{AB}$, at half the value: the potential of B relative to the perimeter is half of the A--B drop, the axis of the channel being the symmetry plane of the injection. The perimeter response to the A--B current, $V_{23}/I_{AB}$ and $V_{12}/I_{AB}$ [Fig.~\ref{fig:sweepab}(c)], is of order $10^{2}$--$10^{3}\,\Omega$ where the contacts are connected to the bulk and drops to below $10^{2}\,\Omega$ where the rings close; it does not vanish, because the Hall potential of the current that crosses the rings still reaches the probes, and it is the quantity that a two-frequency measurement of the Hall voltage at probes 2 and 3 with the A--B excitation would detect.

The interval in which the interior contacts are isolated for an A--B current can be defined by the isolation ratio $\mathcal{R}$ of Table~\ref{tab:regimes} exceeding a threshold; for $\mathcal{R}>10$ it is $\nu_{\rm nom}=1.98$--$2.30$ ($B=5.39$--$\SI{6.27}{T}$, $\Delta B/B=15\%$), for $\mathcal{R}>100$ it is $1.98$--$2.2$ ($11\%$). Its high-field end is fixed to $\pm0.01$ in $\nu_{\rm nom}$ by the disappearance of the incompressible bulk and is independent of $\epsilon$ and of the grid; its low-field end depends on the threshold, and moves to higher fields with increasing $T$ or $\Gamma$ (Sec.~\ref{sec:robust}). The interval lies inside the $R_{xy}$ plateau ($\nu_{\rm nom}=1.98$--$2.375$) and ends before it on the low-field side, which is the saturation of $V_{AB}$ in a subinterval of the plateau observed on the $10\,\mu$m bar~\cite{Kendirlik17}.

\subsection{Robustness of the isolation interval}
\label{sec:robust}

\begin{figure}
\includegraphics[width=\textwidth]{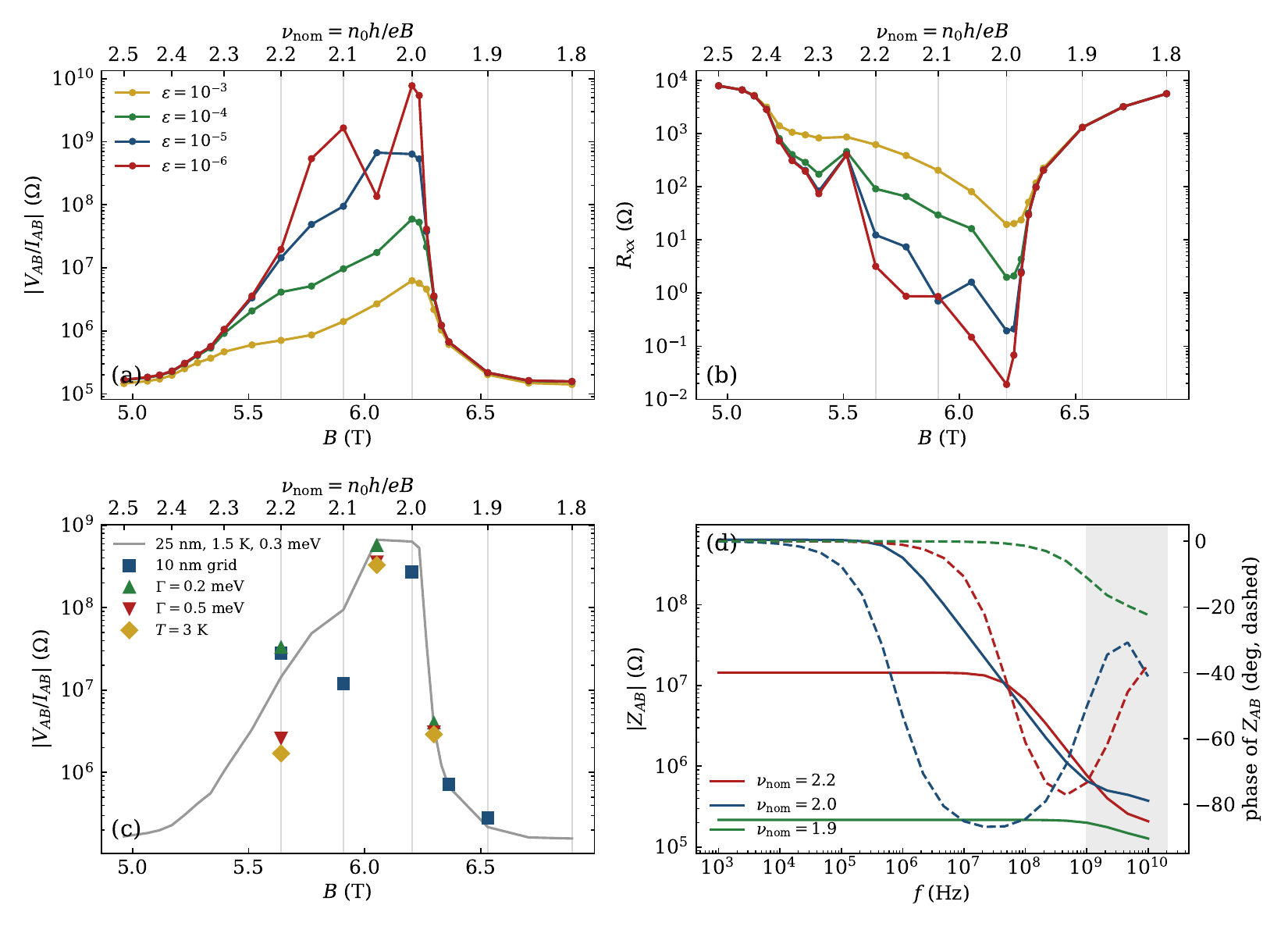}
\caption{Robustness of the isolation interval. (a) Two-terminal resistance $|V_{AB}/I_{AB}|$ and (b) longitudinal resistance $R_{xx}$ (S--D excitation) across the plateau for $\epsilon=10^{-3}$, $10^{-4}$, $10^{-5}$ and $10^{-6}$; outside the isolation interval the curves coincide. (c) $|V_{AB}/I_{AB}|$ for the reference parameters (line) compared with the $10\times25\,{\rm nm}^2$ grid, $\Gamma=0.2$ and $\SI{0.5}{meV}$, and $T=\SI{3}{K}$ at selected fields. (d) Magnitude (solid, left axis) and phase (dashed, right axis) of $Z_{AB}$ for the A--B excitation as a function of frequency at $\nu_{\rm nom}=2.2$, $2.0$ and $1.9$; the shaded region is beyond the validity of the dc conductivity tensor.}
\label{fig:robust}
\end{figure}

Figure~\ref{fig:robust} and Table~\ref{tab:robust} collect the tests of the sensitivity of these results to the regularization, the grid, the Landau-level width and the temperature.

\emph{Residual conductivity.} Figure~\ref{fig:robust}(a,b) shows $R_{AB}^{(AB)}$ and $R_{xx}$ across the plateau for $\epsilon=10^{-3}$ to $10^{-6}$. Outside the isolation interval ($\nu_{\rm nom}\le1.97$ and $\ge2.35$) every resistance is independent of $\epsilon$ to better than 3\%; $R_{xy}$ is quantized to $2\times10^{-4}$ for all $\epsilon\le10^{-4}$. Inside the interval $R_{AB}^{(AB)}\propto1/\epsilon$ and $R_{xx}\propto\epsilon$, i.e. these magnitudes are regularization-limited, but the boundaries of the interval in which they depend on $\epsilon$---which is the operational definition of ``isolated''---do not move. The value of $\epsilon$ is, moreover, not free: it represents the residual conductance of an incompressible region, which in the experiment is finite and measurable. On the $10\,\mu$m bar $V_{AB}$ saturates at $V_{AB}^{\max}=\SI{1.35}{mV}$ for an excitation current $I_{\rm IN}\approx\SI{0.4}{nA}$~\cite{Kendirlik17}, i.e. the two-terminal impedance of the interior contacts rises to $|Z_{AB}|\approx3\times10^{6}\,\Omega$ (about $5\times10^{6}\,\Omega$ once the $\SI{10}{M\ohm}$ series resistor is taken into account as a voltage divider for $V_{\rm IN}=\SI{4}{mV}$), more than an order of magnitude above its value outside the plateaus, at a current of less than a nanoampere. In the model this magnitude is reproduced by $\epsilon\approx10^{-3}$, for which $R_{AB}^{(AB)}=6\times10^{6}\,\Omega$ at the plateau centre and $1.4\times10^{6}\,\Omega$ at $\nu_{\rm nom}=2.1$, against $1.4$--$2\times10^{5}\,\Omega$ outside the plateau [Fig.~\ref{fig:robust}(a)]; the calculated isolation ratio of $10$--$40$ is of the same order as the measured ratio of $V_{AB}^{\max}$ to the baseline of $V_{AB}$ outside the plateaus ($\approx10$--$15$). The saturation value therefore calibrates the residual conductance of the incompressible regions, $\epsilon\,e^2/h\approx4\times10^{-8}\,$S per square, which is the finite thermodynamic density of states in the Landau gap invoked in Ref.~\cite{Kendirlik17}; the smaller value $\epsilon=10^{-5}$ used for the maps corresponds to a cleaner sample and affects only the magnitudes inside the isolation interval. The same holds for the compressible-bulk conductivity: with $\kappa=1$ instead of $0.5$ all resistances outside the plateau scale as $1/\kappa$ and the plateau values change by less than a factor of two; with $\kappa=0.25$ the plateau $R_{xx}$ rises to $50$--$\SI{150}{\ohm}$ and $R_{xy}$ shifts by up to 1\%, because a bulk that conducts only $0.25\nu e^2/h$ competes with the strips, while the ring topology and the isolation interval are unchanged.

\emph{Grid.} On a $10\times25\,{\rm nm}^2$ grid with the conductivity averaged over $\lambda_F$ (Table~\ref{tab:robust}) the number of incompressible regions and their topology are identical to the $25\,$nm results at all five fields, the incompressible areas agree to 10\% and the strip widths to 20\%, and the perimeter resistances outside the plateau agree to 2--5\%. Inside the plateau $R_{xx}$ is larger on the finer grid ($\SI{15}{\ohm}$ instead of $\SI{0.7}{\ohm}$ at $\nu_{\rm nom}=2.1$, $\SI{79}{\ohm}$ instead of $\SI{12}{\ohm}$ at $2.2$) because the finer grid resolves the compressible flanks of the strips; $R_{AB}^{(AB)}$ agrees in order of magnitude [Fig.~\ref{fig:robust}(c)]. The field at which the incompressible bulk disappears is the same on both grids to one sampling step.

\emph{Landau-level width and temperature.} With $\Gamma=0.5$ instead of $\SI{0.3}{meV}$, or $T=3.0$ instead of $\SI{1.5}{K}$, the ring topology is unchanged at all three fields tested, but the isolation at the low-field end ($\nu_{\rm nom}=2.2$) weakens by an order of magnitude ($R_{AB}^{(AB)}$ from $1.4\times10^{7}$ to $2$--$3\times10^{6}\,\Omega$, $R_{xx}$ from $12$ to $40$--$\SI{75}{\ohm}$), whereas at $\nu_{\rm nom}=2.05$ it is unchanged within 40\% [Fig.~\ref{fig:robust}(c)]. The isolation interval therefore shrinks from its low-field side with increasing temperature or disorder broadening, while the plateau centre is unaffected. This is the behaviour observed on the $10\,\mu$m bar, where the saturation interval of $V_{AB}$ shrinks faster with temperature than the plateau~\cite{Kendirlik17}, and it follows from the rings being the narrowest incompressible regions of the sample.

\emph{Frequency.} Figure~\ref{fig:robust}(d) shows $|Z_{AB}|$ and its phase for the A--B excitation between $10^{3}$ and $10^{10}\,$Hz at three fields. Outside the plateau ($\nu_{\rm nom}=1.9$) the impedance is resistive up to $10^{9}\,$Hz. In the isolated regimes the impedance is resistive at low frequency and rolls off with a capacitive phase at the frequency where $\omega C\,{\rm Re}Z\approx1$: since ${\rm Re}Z\propto1/\epsilon$, this corner frequency ($\approx10^{5}\,$Hz at $\nu_{\rm nom}=2.0$ and $\approx10^{7}\,$Hz at $2.2$ for $\epsilon=10^{-5}$) is regularization-dependent, and the phase at a given frequency is not a prediction of the model; what is robust is that the capacitive phase appears only where the contacts are isolated, because $C_{\rm loc}$ vanishes in the incompressible regions, and that it vanishes with the excess real part outside the plateau. The measured phase of $Z_{AB}$ at fixed frequency~\cite{Kendirlik17} is therefore a marker of the isolation interval, while its magnitude is set by the residual conductance of the rings.

\begin{figure}
\includegraphics[width=0.9\textwidth]{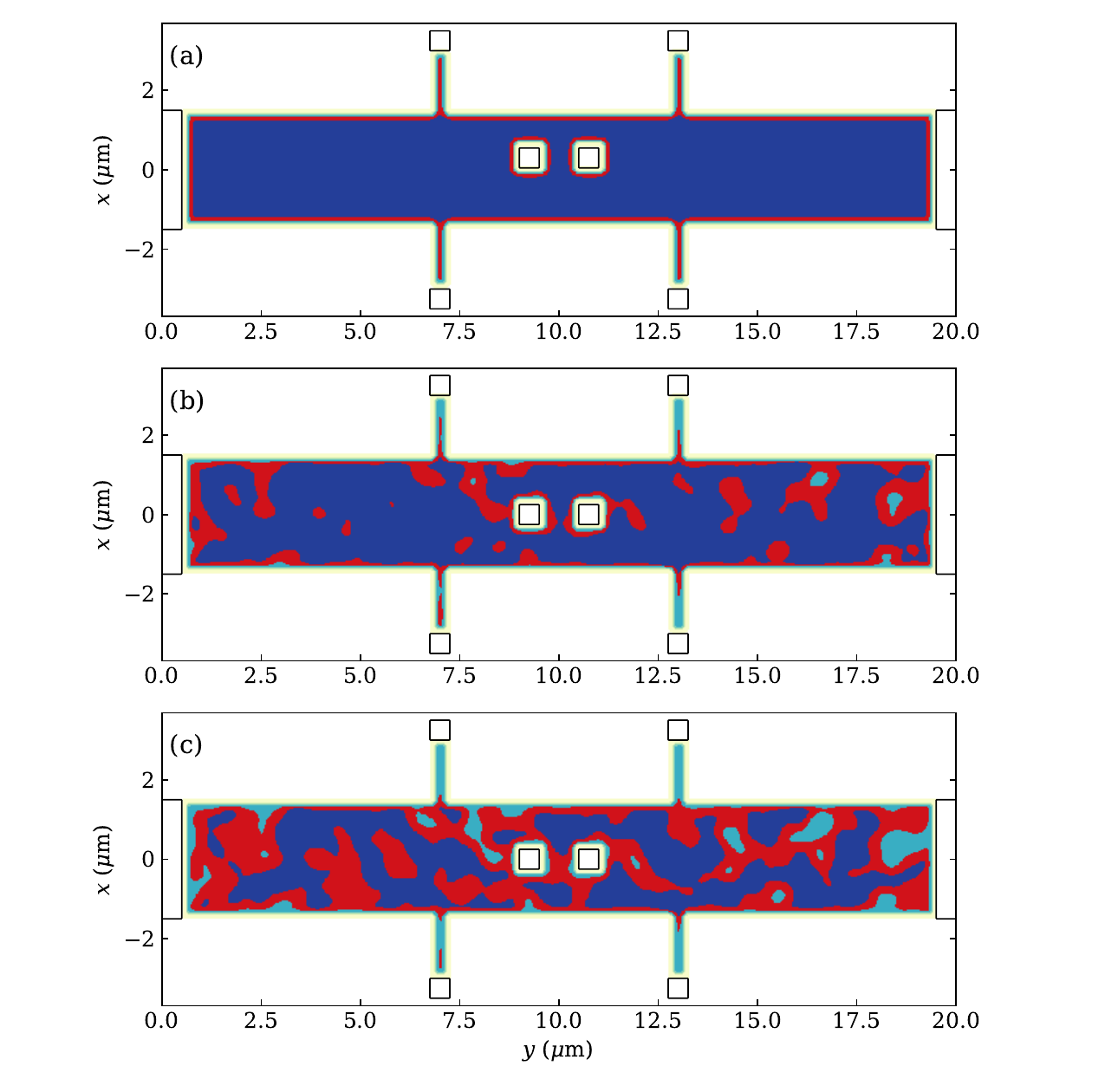}
\caption{Local filling factor (incompressible cells in red) (a) at $\nu_{\rm nom}=2.1$ with the interior contacts displaced by $0.3\,\mu$m from the axis; (b) at $\nu_{\rm nom}=2.1$ and (c) at $2.05$ with a random modulation of the donor density ($3\%$ amplitude, $0.2\,\mu$m correlation length). In all cases the rings around A and B remain closed.}
\label{fig:offaxis}
\end{figure}

\emph{Contact position and disorder.} Two further calculations are shown in Fig.~\ref{fig:offaxis}. With A and B displaced by $0.3\,\mu$m from the axis of the channel, the ring topology, $R_{xy}$, $R_{xx}$ and $R_{AB}^{(AB)}$ are unchanged at $\nu_{\rm nom}=2.1$, $2.0$ and $1.9$, and $R_{B2}/R_{xy}$ moves from $0.50$ to $0.67$ at the plateau centre but stays at $0.50$ in the edge-strip regime, as discussed above. With a random modulation of the donor density of $3\%$ amplitude and $0.2\,\mu$m correlation length, additional closed $\nu=2$ loops appear in the compressible bulk at $\nu_{\rm nom}=2.1$ and $2.2$ and the incompressible bulk at $2.05$ breaks up into islands, but the rings around A and B remain closed: $R_{AB}^{(AB)}$ is $9\times10^{6}$, $6\times10^{7}$ and $2\times10^{8}\,\Omega$ at $\nu_{\rm nom}=2.2$, $2.1$ and $2.05$, within a factor of two of the clean values, $R_{xy}$ stays quantized within $0.6\%$, and $R_{xx}$ rises to $\SI{260}{\ohm}$ at $2.2$ where the strips are narrowest. Long-range fluctuations of this size therefore shift the low-field end of the isolation interval but do not remove it; larger fluctuations, which would reconnect the contacts to the edges through percolating bridges, are discussed in Sec.~\ref{sec:limits}.

\begin{table}
\caption{Grid convergence ($T=\SI{1.5}{K}$, $\Gamma=\SI{0.3}{meV}$, $\epsilon=10^{-5}$): number of incompressible regions $N_{\rm inc}$, incompressible areas (edge strip, rings), width of the edge strip at $y=4\,\mu$m, and resistances on the $25\times25$ and the $10\times25\,{\rm nm}^2$ grid.}
\label{tab:robust}
\begin{indented}
\item[]\begin{tabular}{@{}llllllll}
\br
$\nu_{\rm nom}$ & grid & $N_{\rm inc}$ & areas & $w$ & $R_{xx}$ & $R_{xy}$ & $R_{AB}^{(AB)}$ \\
 & (nm) & & ($\mu$m$^2$) & (nm) & ($\Omega$) & ($\Omega$) & ($\Omega$) \\
\mr
2.2 & 25 & 3 & 2.7, 0.16 & 50 & 12 & 12909 & $1.4\times10^{7}$ \\
2.2 & 10 & 3 & 3.0, 0.17 & 60 & 79 & 12979 & $2.8\times10^{7}$ \\
2.1 & 25 & 3 & 4.7, 0.30 & 100 & 0.7 & 12907 & $9\times10^{7}$ \\
2.1 & 10 & 3 & 4.1, 0.29 & 80 & 15 & 12921 & $1.2\times10^{7}$ \\
2.0 & 25 & 1 & 45.5 & -- & 0.2 & 12906 & $6\times10^{8}$ \\
2.0 & 10 & 1 & 44.6 & -- & 7.8 & 12914 & $3\times10^{8}$ \\
1.95 & 25 & 0 & -- & -- & 206 & 13112 & $6.7\times10^{5}$ \\
1.95 & 10 & 0 & -- & -- & 229 & 13130 & $7.2\times10^{5}$ \\
1.9 & 25 & 0 & -- & -- & 1307 & 13459 & $2.2\times10^{5}$ \\
1.9 & 10 & 0 & -- & -- & 1340 & 13471 & $2.8\times10^{5}$ \\
\br
\end{tabular}
\end{indented}

\end{table}

\subsection{Comparison with the interior-contact experiments}
\label{sec:compare}

The calculation is consistent with the observations on the $10\,\mu$m bar of Ref.~\cite{Kendirlik17} in the following respects: (i) the interior contacts are isolated only in a subinterval of the $\nu=2$ plateau; (ii) this subinterval is offset towards the high-field side of the plateau---here it ends at $\nu_{\rm nom}=1.98$ together with the plateau but starts at $2.2$--$2.3$ while the plateau extends to $2.375$; (iii) it shrinks from its low-field side with temperature while the plateau is unaffected; and (iv) the transition at its high-field end is sharp (here $\Delta B/B\approx0.5\%$) because it is caused by the disappearance of the incompressible bulk, whereas the low-field end is gradual because the rings narrow continuously. In magnitude, the measured saturation of $V_{AB}$ corresponds to a two-terminal impedance of the interior contacts of $3$--$5\times10^{6}\,\Omega$ at $I_{\rm IN}\approx\SI{0.4}{nA}$, which the model reproduces with a residual conductance $\epsilon\approx10^{-3}\,e^2/h$ of the incompressible regions (Sec.~\ref{sec:robust}); the fact that the impedance becomes this large at such a small current is the direct evidence that the interior contacts are decoupled by an insulating region, not by a resistive bulk, and it is reproduced by the model only when the rings are closed or the bulk is incompressible. Beyond this, the comparison is qualitative: the device widths (3 and $10\,\mu$m), the contact sizes and separations differ, and the parameters $\Gamma$, $l_{\rm pin}$, $\epsilon$ and $\kappa$ are not fitted to the experiment. The scaling from the narrow to the wide bar follows from the strip widths being set by the electrostatics and not by $W$: the fraction of the plateau in which the rings are wider than $\lambda_F$, and hence the isolation interval in $\Delta B/B$, is expected to be similar for the two widths, whereas the absolute size of the incompressible-bulk regime shrinks with $W$ because a wider bulk requires a more homogeneous density to be incompressible everywhere.

The transition from edge- to bulk-currents with increasing $B$ that is demonstrated experimentally in Ref.~\cite{Sirt25} on an interior-contact device is the same sequence as Figs.~\ref{fig:cur22}$\to$\ref{fig:cur20}: at the low-field side of the plateau the current is carried by the edge strips and the interior contact is decoupled by its ring, at the high-field side it is carried by the incompressible bulk. The impedance surge of the interior contacts in Fig.~\ref{fig:sweepab} occurs at the field at which the current-carrying incompressible region reaches the contacts, first as a ring and then as the whole bulk, and the abrupt end of the surge at $\nu_{\rm nom}=1.98$ marks the field beyond which the bulk becomes compressible and the current returns to the bulk.

One aspect of the interpretation given in Ref.~\cite{Sirt25} is not supported by the present calculation, namely the assignment of a chirality to the currents in the two edge strips in the sense of the Landauer--B\"uttiker edge channels, which carry the current in opposite directions on the two sides of the bar. In the local Ohm's law with the self-consistent conductivity tensor the two edge strips carry the imposed current in the \emph{same} direction, from S to D, and share it equally: integrating $j_y$ over the two halves of the channel at $y=4\,\mu$m and at $y=10\,\mu$m gives $0.50\,I$ on each side to within 2\% at $\nu_{\rm nom}=2.1$ and $2.2$ [Figs.~\ref{fig:cur21}(c) and \ref{fig:cur22}(c)], as it does for the incompressible bulk at $\nu_{\rm nom}=2.0$. What is chiral is the Hall potential, which is high on one edge and low on the other, not the net current; the direction of the current in a strip is fixed by $\bm j=\hat\sigma\bm E$ with $\bm E$ the gradient of the electrochemical potential along the strip, which has the same sign on both edges because both connect the source to the drain~\cite{Siddiki04,Guven03}. The distinction matters for the interior contacts: a contact isolated by its ring sits at the potential of the flat bulk between the two strips, half-way between the two edge potentials, which is why $R_{B2}$ approaches $h/4e^2$ independently of the side of the bar on which the interior contact is displaced (Sec.~\ref{sec:robust}); in a chiral-edge-channel picture the potential of an interior contact would be set by the channel it equilibrates with and would take one of the two edge values. The comparison should not be pressed too far: the local Ohm's law describes a diffusive, dissipative medium and defines the current density at every point, whereas the Landauer--B\"uttiker chirality is a statement about the direction in which a ballistic channel carries \emph{excess} occupation between reservoirs, and the two descriptions agree on the total current and the terminal voltages. What the calculation excludes is only the stronger, real-space reading of the chirality, in which the physical current in one edge region flows opposite to that in the other; in the screening theory the net current in both edge strips flows from source to drain, and the interior contacts probe the potential of the region between them.

The interior contacts add a piece of information that the perimeter probes cannot supply. On the high-field side of the plateau they are part of a resistive current path and $V_{AB}$ scales with $I$ and with the longitudinal resistivity. At the plateau centre they float in a dissipationless bulk; for an S--D current $V_{AB}$ probes the local Hall potential, whereas for an A--B current the impedance is large and, at finite frequency, capacitive. On the low-field side the contacts are decoupled from the current by closed incompressible rings; in a strictly dissipationless model their potentials are undefined, and in practice they are set by the small residual conductance of the rings (thermal activation, tunnelling across the strip, localized states in the Landau-level tails, or scattering at the ring corners), which is exactly what fixes the finite saturation value $V_{AB}^{\max}\propto1/D_T(E_F)$ in the experiment~\cite{Kendirlik17} and what the parameter $\epsilon$ stands for in the model. The complementary experiment, in which a current is driven between A and B and the Hall voltage $V_H$ is read at probes 2 and 3, is described by $V_{23}/I_{AB}$ in Fig.~\ref{fig:sweepab}(c): a response of a few hundred ohm where the bulk is compressible, and below $10^{2}\,\Omega$ once the rings close. In the experiment the two excitations were applied at different frequencies, which separates the two linear-response coefficients in the detection~\cite{Kendirlik17}, while at large excitation the edge strips themselves break down~\cite{Sirt23}.

\subsection{Role of the two-dimensional self-consistency}

Several features of the results depend explicitly on solving the electrostatics in two dimensions for the actual device rather than for a translation-invariant strip or a periodic cell. First, the rings around the interior contacts are closed and are not connected to the edge strips at any of the fields studied, even though the contact halos and the edge depletion are separated by less than $1\,\mu$m at $B=0$. A one-dimensional treatment of the cross-section through the contacts would produce four $\nu=2$ strips at each $y$ but could not decide whether the inner pair closes around the contacts or merges with the outer pair along $y$; here the closure follows from the Poisson solution, as it did for the wider bars of Refs.~\cite{Yildiz14,Sirt23}. Second, the current distribution at the plateau centre and the leakage near the probe arms at $\nu_{\rm nom}=2.2$ are determined by the concave corners of the geometry, at which the incompressible region has to bend; these are absent in a translation-invariant model and in the periodic cells used earlier. The probe arms themselves, which are usually ignored in strip calculations, are seen to host $\nu=2$ strips continuous with those of the channel, which justifies treating the perimeter probes as ideal voltage terminals in the plateau; where this continuity is broken by density fluctuations in front of the probes, the quantization is expected to suffer~\cite{Eksi10}. Third, the source and drain are here depleted contacts of finite conductivity rather than the ideal injectors of Ref.~\cite{Sirt23}, so that within the model the current enters the edge strips through the compressible rims at the corners, the hot-spot configuration of Refs.~\cite{Eksi10,Ahlswede01}.

\subsection{Limitations}
\label{sec:limits}

The calculation neglects spin splitting (Sec.~\ref{sec:tfpa}) and uses a local conductivity model that requires the incompressible strips to be wider than $\lambda_F$~\cite{Siddiki04}. The latter condition holds for the edge strips at $\nu_{\rm nom}=2.1$ and $2.2$ and for the rings at $2.1$; at $2.2$ the rings are about $\lambda_F$ wide and their insulating character is an upper bound. The model contains no tunnelling or hopping across an incompressible region; all such processes are represented by the residual $\epsilon$, which is why the magnitudes inside the isolation interval are not predictions (Sec.~\ref{sec:robust}). The omission matters where the incompressible regions are narrowest. An electron crossing a strip of width $w$ has to be displaced by $w$ in the guiding-centre coordinate, so that the transmission is of order $\exp(-w^2/4\ell_B^2)$~\cite{SiddikiMarquardt07,Kendirlik13}: at $B=\SI{5.6}{T}$ ($\ell_B=\SI{10.8}{nm}$) this is $5\times10^{-3}$ for the $\SI{50}{nm}$ rings of $\nu_{\rm nom}=2.2$, $4\times10^{-4}$ for $\SI{60}{nm}$ and $5\times10^{-10}$ for the $\SI{100}{nm}$ strips of $\nu_{\rm nom}=2.1$, while thermal activation across the cyclotron gap ($\hbar\omega_c=\SI{9.8}{meV}$) is entirely negligible at $\SI{1.5}{K}$. Tunnelling therefore reduces the isolation only at the low-field boundary of the interval, where it acts in the same direction as the thermal and disorder effects of Sec.~\ref{sec:robust}: it is an additional mechanism by which the rings become leaky before the edge strips do, and it makes the calculated low-field end of the isolation interval an upper bound. It is suggestive that the width at which this transmission equals the $\epsilon\approx10^{-3}$ calibrated from the experiment, $w\approx\SI{57}{nm}$, is the ring width at $\nu_{\rm nom}=2.2$; for the wider strips and for the incompressible bulk the residual conductance must instead come from the localized states in the Landau-level tails, which is the finite thermodynamic density of states that the measured $V_{AB}^{\max}$ determines~\cite{Kendirlik17}. A constant $\epsilon$ cannot capture both limits, and a width-dependent residual conductance would sharpen the low-field end of the interval further. Disorder enters only through the Landau-level broadening and, in one test, through a $3\%$ modulation of the donor density. Long-range fluctuations of the density on the scale of the spacer ($\SI{45}{nm}$), of a few per cent, would open the rings first at $\nu_{\rm nom}=2.2$, where they are narrowest, and shift the low-field end of the isolation interval upward in field; fluctuations on the scale of the ring-to-edge distance ($0.7\,\mu$m) would create closed $\nu=2$ loops in the bulk that can either reconnect the contact islands to the edges through percolating bridges, or extend the incompressible-bulk regime, as found for disordered bars and Corbino discs in Refs.~\cite{Yildiz14,Sirt23,Gulebaglan12}; the isolated regime survives as long as the rings themselves stay closed. The current is treated in linear response. In the local model an incompressible strip breaks down when the Hall voltage across it, $I\rho_H$, approaches the Landau gap, i.e. at $I_c\approx\nu e\,\hbar\omega_c/h\approx0.8\,\mu$A per strip at $\SI{6}{T}$~\cite{Siddiki04,Gerhardts13}; the imposed current of the present calculation, $I=\SI{1}{nA}$, is thus almost three orders of magnitude below $I_c$, and the A--B current of the experiment, $I_{\rm IN}\approx\SI{0.4}{nA}$~\cite{Kendirlik17}, more than three, so that the rings are not broken by the excitation but the compressible islands around the contacts are heated~\cite{Akera00,Akera01,Kanamaru06,Yurdasan13}; since the rings are only $2$--$4\,\lambda_F$ wide, the local estimate of $I_c$ is an upper bound. Current-induced asymmetries of the strips~\cite{Siddiki09,Siddiki10,Gerhardts13} would primarily affect the widths of the rings. Finally, the contacts are treated as ideal equipotential columns with a smooth depletion halo; the inhomogeneous NiGe/Au grain structure of real alloyed contacts~\cite{Eksi10,Dahlem10} would roughen the halo and shift the field at which the rings close.

\section{Conclusion}
\label{sec:conclusion}

We have carried out fully two-dimensional self-consistent screening calculations for a $3\,\mu$m wide six-terminal Hall bar containing two interior contacts on the axis of the channel, in the geometry of the interior-contact experiment of Kendirlik \emph{et al.}~\cite{Kendirlik17}. The zero-field density shows that the depletion halos of the contacts remain separated from each other and from the channel edges. Around the $\nu=2$ plateau the current distribution passes through three regimes: bulk current in a compressible sample on the high-field side of the plateau, bulk Hall current in a fully incompressible sample at the plateau centre, and edge-strip current with a compressible bulk on its low-field side. In the last regime each interior contact is enclosed by a closed incompressible ring that carries no current and isolates the contact from the current-carrying edges. The interior contacts are consequently predicted to switch from being galvanically connected to the source--drain circuit, to floating in the Hall potential, to being isolated, as the field is lowered through the plateau; the isolation interval, $\nu_{\rm nom}=1.98$--$2.3$ for the parameters used, is bounded sharply at its high-field end by the disappearance of the incompressible bulk and gradually at its low-field end by the narrowing of the rings, which are the first incompressible regions to become leaky as the field is lowered or the temperature raised. The topology of the incompressible regions and the boundaries of this interval are independent of the regularization of the local conductivity, of the grid and of moderate disorder, whereas the magnitude of the isolation resistance is not. The voltage between an interior contact and a perimeter probe approaches one half of the quantized Hall resistance across the whole plateau in the symmetric geometry. This is the sequence inferred from the saturation of $V_{AB}$ in a subinterval of the plateau, its offset from the plateau centre and its stronger temperature dependence on the $10\,\mu$m bar~\cite{Kendirlik17}, and it is consistent with the edge-to-bulk transition of the current observed in Ref.~\cite{Sirt25}, now obtained for a narrow bar with the actual six-terminal geometry. Both $V_{AB}$ for a source--drain current and the Hall response to an A--B current should reflect this switching and provide a transport signature of the compressibility of the bulk that complements scanning-probe imaging.

\ack
The author thanks the colleagues involved in the interior-contact experiments for discussions.

\end{document}